\documentclass{article}

\usepackage{arxiv}

\usepackage[utf8]{inputenc} 
\usepackage[T1]{fontenc}    
\usepackage{hyperref}       
\usepackage{url}            
\usepackage{booktabs}       
\usepackage{amsfonts}       
\usepackage{nicefrac}       
\usepackage{microtype}      
\usepackage{lipsum}
\usepackage{graphicx}
\usepackage{amsmath}
\graphicspath{ {./images/} }

\title{Convergence and Inequality in Research Globalization}

\author{
Saurabh Mishra \\
  Institute for Human-Centered Artificial Intelligence (HAI)\\
  Stanford University, Stanford, California, USA\\
  \texttt{saurabh.mishra@stanford.edu} \\
   \And
Kuansan Wang \\
  Microsoft Research\\
  Microsoft Corp., Redmond Washington, USA\\
  \texttt{kuansan.wang@microsoft.com} \\

}

\begin{document}
\maketitle
\begin{abstract}
The catch-up effect and the Matthew effect offer opposing characterizations of globalization: the former predicts an eventual convergence as the poor can grow faster than the rich due to free exchanges of complementary resources, while the latter, a deepening inequality between the rich and the poor. To understand these effects on the globalization of research, we conduct an in-depth study based on scholarly and patent publications covering STEM research from 218 countries/regions over the past four decades, covering more than 55 million scholarly articles and 1.7 billion citations. Unique to this investigation is the simultaneous examination of both the research output and its impact in the same data set, using a novel machine learning based measure, called saliency, to mitigate the intrinsic biases in quantifying the research impact. The results show that the two effects are in fact co-occurring: there are clear indications of convergence among the high income and upper middle income countries across the STEM fields, but a widening gap is developing that segregates the lower middle and low income regions from the higher income regions. Furthermore, the rate of convergence varies notably among the STEM sub-fields, with the highly strategic area of Artificial Intelligence (AI) sandwiched between fields such as Medicine and Materials Science that occupy the opposite ends of the spectrum. The data support the argument that a leading explanation of the Matthew effect, namely, the preferential attachment theory, can actually foster the catch-up effect when organizations from lower income countries forge substantial research collaborations with those already dominant. The data resoundingly show such collaborations benefit all parties involved, and a case of role reversal can be seen in the Materials Science field where the most advanced signs of convergence are observed.
\end{abstract}



\section{Introduction}\label{sec:intro}

The exchange of people and ideas is the cornerstone of globalization. Research globalization implies collaboration between countries. Multidisciplinary scientists have extolled various forms of globalization, including economic, financial, cultural, or political, but technology globalization, which is the dominant driver of globalization in the $21^{st}$ century, is less studied. Technology (or scientific) globalization creates opportunities for national competitiveness and forms the backbone of long-run economic growth and development \cite{aghion2009science, mokyr2018building, verspagen2006innovation}. Incomplete or incorrect measures to track technology globalization could lead to misleading conclusions for global and national science policy. 

The central theme of this paper is to explore whether poor countries are catching up with rich countries in the realm of scientific research or, alternatively, the gap in technology research output and impact is widening between rich and poor countries. In natural and social sciences, these trends are referred to as convergence or divergence. Convergence refers to coming together, while divergence means moving apart or an increase in inequality \cite{macarthur1967limiting, durlauf1996convergence, frost2020matthew}. Convergence implies that the lesser will catch up by growing faster than the greater in terms of research output and impact. Globally, there is a growing shift from so called “national system(s) of innovation” to a series of globally connected networked systems of knowledge creation \cite{simon2012globalization}. Technologies such as Artificial Intelligence (AI) could qualify as general-purpose technologies (GPTs) that have many applications across a wide spectrum of uses in production and research. Most encouragingly, recent advancements there are propelled by open data sharing, open source software and other open science principles that considerably lower the access barriers \cite{wang2019opportunities}. There is a growing list of indicators, including composite indices that are used to guide global and national science (more generally) and AI policy (more specifically) \cite{perrault2019ai, grupp2004indicators, haynes2019global, aioecd}. Many of them lend support to convergence as a necessary outcome, where it can be observed that the dominant are becoming less so. On the other hand, it is also widely observed that inequality, characterized by the Matthew Effect \cite{merton1968matthew}, is deeply entrenched in the global research ecosystem, confirmed again in a recent study that shows top-cited scientists, often affiliated with high-ranking institutions in the western world, are receiving ever increasing citations \cite{nielsen2021global}. Similarly for AI, deep learning's unanticipated rise appears to have resulted in divergence, where compute divide between large firms and non-elite universities increases concerns of "de-democratization" of knowledge production \cite{ahmed2020democratization}. These observations motivate this study to view convergence and divergence as yin and yang, namely, the twin effects of research globalization so that we can avoid the pitfall of missing half of the puzzle by looking for indicators of only convergence or only inequality.

The unique contribution of this paper is thus to provide a more complete view of research globalization by accounting for research output, impact, and collaboration to answer fundamental questions in research globalization. Is there hope for developing countries to catch up with advanced economies in research output and impact? Countries can generate a large amount of research output with relative impact not increasing in a commensurate manner. How do countries get out of this trap? What is the role of science policy, especially for developing countries to catch up with rich countries? Our data suggest that the example of China could provide an alternative path for developing countries. The output and impact of technology research from China in the 1980's was at par with other developing countries such as India. However, in recent years, China's technology research has been catching up with that of rich countries, first in output and subsequently, in impact. If developing countries build research collaborations with advanced economies as China has been doing, could that lead to global research convergence?

The study employs a quantitative and data-driven approach, following the similar methodologies reported in the literature for general research \cite{zhou2006the, xie2014china, leydesdorff2014european, dong2017a, tollefson2018china, xie2019bigger, nsf2019} and specifically for AI \cite{perrault2019ai, haynes2019global, aioecd}. We take into account the global corpora of journal publications, conference papers, and patents to answer three key questions. First, is there global convergence in research output and impact across sub-fields of technology research? Second, what are the dynamics of convergence (or divergence) between comparable income groups? Third and perhaps most novel, which country pairs have emerged as central, field-specific collaborators and what are the science policy implications to avoid the output-impact trap?

The results show that convergence is taking place, but only in the case of Upper-Middle Income countries (UMCs) catching up with High-Income countries (HICs), following the World Bank's 2020 income categorization. Similar to \cite{abbasi2012betweenness}, we find that, on average, rich countries are becoming less dominant as they are losing their share in world research output and impact. However, if we separate Middle-Income countries (MICs) into UMCs and Lower-Middle-Income countries (LMCs), the results are mixed. These two groups started out with a similar low base of output and impact but, in recent years, the gap between UMCs and LMCs is widening, supporting evidence for divergence. The results vary by field --- for example, China, an UMC, is catching up in fields such as AI or Materials Science but not Medicine. Low-income countries (LICs) are diverging and do not show signs of catching up. LICs are not catching up with LMCs, and LMCs are not catching up with UMCs. The data nevertheless supports that neither convergence nor divergence is a preordained outcome of globalization. The contrasting trends between UMCs and LMCs appear to support a formula for catching up by substantial research collaboration. The results about research collaboration provide hope for science policy, especially in developing countries. For example, STEM and AI research in China is driven by research culture \cite{shi2010china} and policy incentives to formulate R\&D output milestones \cite{cao2013reforming, yang2016policy, cao2006china}. It is key to recognize that China was able to reach from a small base in the 1990's to take over market share in STEM and AI, and collaboration between China and G-7 countries has been identified as critical to global R\&D growth \cite{he2009international, han2018china}. We show evidence that collaboration between the US and China in AI and STEM fields offers hope for developing countries to increase output and impact. Conscious policy efforts to build field-specific research collaboration can help developing countries increase their scientific contribution and impact.
  
The rest of the paper is divided as follows. Section 2 presents details about the data and  methodology for the quantitative measure used in this study. Section 3 presents the results segmenting metrics of research output, impact, collaboration, and saliency comparing world regions between STEM and AI fields of study. Section 4 provides discussion on policy implications with areas of future research.

\section{Data and Method}
\label{sec:methodology}

Following the well-established practice first proposed by Price \cite{price1965networks}, we measure research output and impact through two proxy activities, namely, the participation and the recognition received in the global scholarly communications. In a nutshell, this study quantifies research output by the number of scholarly articles published, and research impact, by the citations they have and likely to receive. There are known issues in using publication and citation measures, and plenty of other options have been proposed in the scientometric, bibliometric, or science of science literature, with the most relevant ones discussed below. However, we note that a widely accepted measurement for convergence has yet to emerge, and the causality between globalization and convergence is tricky to prove. Accordingly, we approach the question of convergence by examining whether the historical elite dominance is lessening (or the relative share share of global research output and impact is increasing for developing countries). For this purpose, we contend it adequate to base the analyses on the time honored metrics with straightforward enhancements as described below.

\subsection{Data}\label{sec:data}
Similar to~\cite{dong2017a}, the study is conducted with Microsoft Academic Graph (MAG) but on the newer version dated on August~7, 2020 and validated against the version dated October~29, 2020. These two versions of MAG allow multiple affiliations to be associated with each author, in contrast to the data used previously where only the primary author affiliation was available. The new versions of MAG also provide the geolocation and ISO-3166 region code for each affiliation, with which we classify the world into the following regions based on 2020 World Bank income codes (included in the data in the Supplementary Material). Within this study, the income codes used are HIC, UMC, LMC, and LIC, which correspond to World Bank's classification of high income, upper middle income, lower middle income, and low income countries, respectively. No countries simultaneously belong to more than one categories. The numbers of countries/regions in these categories are shown in Table~\ref{tab:my_label}.

Since the objective here is to gain deeper understanding of the global trends in recent decades that coincide with the accelerating globalization propelled by the internet and digital transformation, we focus on the time horizon of year 1980 (inclusive) through 2020 (exclusive). To avoid the data quality issues, this study reports only the results in the STEM areas that better align with MAG’s selection preference of including internationally oriented articles, typically written in English, than the fields in the humanity area where English publications do not provide an adequate characterization of the state of research. The STEM areas, corresponding to the topmost 12 MAG fields of study, are Medicine, Biology, Chemistry, Computer Science, Engineering, Material Science, Physics, Psychology, Environmental Science, Geography, Geology, and Math. The analytical scripts provided in the supplementary material, though, can be extended to broader fields and year range in a straightforward manner. In total, this study inspects 56,084,649 scholarly articles and their 1,710,566,975 citations as recorded in MAG.

There are uncertainties associated with the global corpora. For example, the class imbalance in limited number of low-income countries (LIC's) relative to UMC's or HIC's. Further, although although MAG strives to be comprehensive, it is by no means perfect for this study in the following aspects. First and foremost, MAG by design includes publications that are international oriented \cite{wang2019a}, which leads to a lower representation of articles written in languages other than English. The data set thus has a bias against non-English research and can under-count research output of countries where local language publications are not always translated into equivalent English versions, for example, Japan (a HIC), Russia, Brazil and China (all UMCs). This study is only focused on English-language research output. Secondly, the affiliation coverage is uneven, especially for older publications, as many of its data sources do not report this information. Although MAG uses PDF parsing to extract such information \cite{wang2019a}, we notice such results are not error free. Finally, MAG's affiliation attribution may be not fine grain enough, especially for work produced by multi-national industrial institutions for which affiliation are attributed to the countries/regions where their headquarters' reside. For example, publications from Microsoft Research Asia, India, and Cambridge labs are all counted towards the US, rather than China, India, and UK, respectively. This attribution policy under-represents the research activities of their host countries and inflate the international collaboration (Sec.~\ref{sec:collaboration}) when these researchers work with nearby universities. An alternative is to use the Global Research Identifier Database (GRID), a budding community-wide grass root effort to annotate finer grained information of affiliations led by Digital Science\footnote{See https://digital-science.com/products/grid for more details}. Earlier investigations show the GRID code in MAG does not lead to more accurate results than ISO-3166 code, but the analytical software included in the Supplementary Materials can be easily adapted for GRID should the data quality improve.
	
\begin{table}[t]
    \centering
    \begin{tabular}{|l|c|}
    \hline\hline
        Category & Number of Countries/Regions \\
        \hline\hline
       High income country (HIC)  &  83 \\
       Upper middle income country (UMC) & 56 \\
       Lower middle income country (LMC) & 50 \\
       Low income country (LIC) & 29 \\
       \hline\hline
       Total & 218 \\
       \hline\hline
    \end{tabular}
    \caption{World Bank's classification of the 218 countries/regions and their sizes}
    \label{tab:my_label}
\end{table}

\subsection{Research Output}\label{sec:output}

The annual research output is characterized by the number of articles published in the given year of a given field with at least one author affiliated with an institution in the target region. There is no weighting on the authorship, e.g., one author affiliation from region A and ten from region B leads to the equal attribution to either region. Similarly, there is no weighting or fractional counting based on the order in which an author appears in the author list. In the case of an author with multiple affiliations, the order of affiliations is also ignored. Note that an article with affiliations from multiple regions or classified as more than one field is counted multiple times. The sum of the article counts from individual regions and fields therefore exceeds the total number of articles published.

\subsection{Research Impact}\label{sec:impact}

Three indices are considered in this study to assess research impacts: citation counts, the saliency, and the saliency ratio. 

\subsubsection{Citation Count} 
The total citations received by articles in a given year with authors from a world region are aggregated from MAG’s records as a conventional baseline to characterize the research impact for the region. Similar to the publication counts above, regions received citation attributed uniformly in the sense that no author weighting is performed and all affiliations of a single author are all treated equally when tabulating the citations. 

Citation count described above is a commonly used metric of research impact, and can either be used alone for measuring individual articles or in aggregations such as in Journal Impact Factor (JIF) \cite{cagan2013san} for measuring journals, h-index for authors \cite{waltman2012inconsistency}, or share of highly cited articles for countries/regions \cite{nsf2019}. In this study, we also report citation counts aggregated at the country or regional levels as a baseline measure of research impact. Known issues for citation counts, however, are plenty. First, citations exhibit a strong age bias against newer articles because it takes time for a new publication to receive its due recognition. The estimation of a lag ranging from 7 to 10 years, first made by Price in 1965 \cite{price1965networks}, has withstood the test of time. Citation count is therefore a lagging indicator of research impact that is not suitable to discover recent trends, as can be seen in the results of this study later. 

Secondly, citation behaviors are known to vary significantly from one field to another \cite{waltman2019field}, and lumping the analyses of different fields into a general statement can over-represent the fields where the community favors more references in the research articles. This challenge has led to the development of field-normalized or weighted modifications on citation counts, such as Mean Normalized Citation Score (MNCS), also known as the crown indicator \cite{waltman2011towards}, Relative Citation Ratio (RCR) adopted by National Institute of Health in PubMed \cite{hutchins2016relative}, and the Field Weighted Citation Index (FWCI) by Elsevier that has been empirically shown to be consistent with the MNCS and RCR \cite{purkayastha2019comparison}. However, when citation counts of individual articles are normalized by different factors, they pose a mathematical challenge to be aggregated into an author or country level measure mainly because $E[X/Y] \neq E[X]/E[Y]$, namely, an order reversal in normalization and aggregation leads to very different measurement. Despite the successful applications, the normalization methods are often more heuristically than mathematically motivated that may lead to unexpected and inexplicable outcomes. For instance, FWCI has been used in numerous studies to guide policy and identify top countries in AI research \cite{perrault2019ai, haynes2019global, tollefson2018china, xie2014china, zhou2006the}. However, in these instances, the US is ranked behind Georgia, Hong Kong, Switzerland, and Singapore, with Qatar, Palestine, Israel, and Belarus following closely ahead of European Union (EU) countries (Figure A1.5, \cite{perrault2019ai}). These results defy the shared understanding of the AI community, and cannot be collaborated in this study. Still, as studies like these are frequently consulted in making consequential policy or funding decisions, the importance of quantitative measurements based on a solid mathematical foundation cannot be over-emphasized. 

Furthermore, citations are often treated equal without considering the topical relevance and the reputation of citing sources. This measurement weakness, often exacerbated by the research reward system \cite{chapman2019games}, has led to schemes of “gaming the system” where abusive behaviors, such as citation cartels \cite{fister2016toward, perez2019network} and coercive citations \cite{herteliu2017quantitative, wilhite2012coercive, haley2017inauspicious}, have become an ever more concerning issue for research integrity and assessments. Research into better measurements of impact have been active \cite{fortunato2018science, aksnes2019citations, zhou2017connecting, liao2017ranking, waltman2016a, ahmadpoor2019decoding}, though widely accepted alternatives have yet to emerge.

\subsubsection{Saliency Measure}
To address these issues, the saliency measure in MAG is used as a supplementary metric for research impact in this study. As described in \cite{wang2019a}, the gaming of the citation counts is treated as a link spam detection problem \cite{benczur2005spamrank, gyongyi2005link, zhou2007transductive}, and the known effective method, namely, using the eigenvalue centrality of the citation graph, is employed in the saliency design to combat this problem. Popularized by Google's PageRank and widely experimented in bibliometric studies \cite{waltman2016a}, the eigenvalue centrality uses the number of references made to the cited articles as an indication to the relevance of the citation, and citations from more reputable sources carry more weights. Both are desired properties in alleviating the effects from the illicit citation behaviors. 

Additionally, the saliency measure utilizes the time stamp on each citation to capture the temporal dynamics of the eigenvalue centrality to avoid biases against newer articles. Rather than simply counting citations that have already occurred, the saliency measure is designed to estimate the eigenvalue centrality of the citation graph 5 years into the future, leveraging the remarkable prediction power of the reinforcement learning algorithm that is convincingly demonstrated in DeepMind's AlphaZero \cite{silver2017mastering} in anticipating the opponent's future moves in the game of Go. Whereas in a game setting the learner can be programmed to play against itself to harvest the sufficient amount of training data, MAG has over two centuries worth of citation records for the learner to travel back in time and learn to predict. The outcome is a measure that characterizes the {\em likelihood} of an article to be highly cited in the next 5 years. As research is constantly built upon accumulated knowledge and research articles tend to cite up-to-date information, empirical evidence \cite{wang2019a} suggests the saliency measure can effectively identify recent articles as having higher impact potential than older ones, an observation that can also be made in this study (Sec.~\ref{sec:results}).

Finally, the probabilistic nature of an eigenvalue centrality also provides a mathematically rigorous framework to aggregate article-level impact from various fields into a measure of a higher level. To be more specific, let $s(c,v,t,f,a,o)$ denote the saliency of the article $c$ published at venue $v$ on the publication date $t$ about a topic $f$ by an author $a$ affiliated with an organization $o$. The saliency of a region $W$ on an aggregate field $F$ at time $t$ can be obtained by computing the marginal probability distribution from the saliency as
\begin{equation}
s(F, W, t) = \sum_{f \in F} \sum_{o \in W} \sum_{v} \sum_{a} \sum_{c}  s(c, v, t, f, a, o) \label{eq:saliency}
\end{equation}
Similarly, the saliency of an author or an venue can be computed by summing the saliencies of all articles by the author or in the venue. Putting it all together, an article must receive sustained references from high saliency articles by high saliency authors published in high saliency venues. Citation cartels with unknown authors using articles published in low quality journals, for example, may distort the citation count of a target article more easily than manipulate its saliency.

As eigenvalue centrality is also used in many economic studies, most notable the Nobel winning input-output model, it is not surprising that the properties of saliency can be understood in the same manner. Particularly, saliency could be viewed to bibliometric analysis akin to price indices, i.e., inflation in economics. For example, a 1-2\% inflation over one or two-year horizon could be barely noticeable but over a long period such as a decade, could erode 20\% of the cash value. The time value of money decreases the value of a currency over time. In similar manner, citation metrics are fraught with biases stemming from authorship, fields of study, concepts, entities involved (sponsoring agency, university, etc.), or time-lag of accruing citation. These biases could erode the time value of scientific contribution. In the same manner inflationary bias are caused \cite{driffill1988macroeconomic, seiler2020weighting}, the life cycle of research could be biased by many factors. Saliency accounts for biases caused to citation metrics by accounting for temporal, field-wise, geographic variations through an explainable probabilistic measure to normalize research impact. 

\subsubsection{Saliency Ratio}
In this work we study the trend of convergence by examining the signs that suggest the elite dominance is attenuated with a specified condition, such as within in time frame or within a region. While the saliency measure, even the marginalized one of~\ref{eq:saliency}, gives an absolute measurement agnostic to the underlying condition, its probabilistic nature affords a well motivated and mathematically rigorous processing, namely, the conditional probability, to quantify the performance of an entity ``relative'' to its peers under the condition. Such relative measure is quite intuitive as it describes the percentage share of the entity. For instance, to study how dominant a country $C$ in region $W$ is in generating impact for a research area $F$ at time $t$, we quantify the degree of dominance as the conditional probability
\begin{equation}
R(F, C\subseteq W, t) = s(F, C, t|W) = \frac{s(F,C,t)}{s(F,W,t)} \label{eq:saliencyRatio} 
\end{equation}
where $s(F,C,t)$ can be obtained by replacing the summation over $o\in W$ with $o\in C$. 

As the conditional probability, by definition, assumes the form of a ratio between the intersection event versus the condition event, we call the conditional saliency the {\em saliency ratio} in the following.

\subsection{Research Collaboration}\label{sec:collaboration}

The trends in the world regions are studied in terms of their individual research output and impact. Additionally, as the previous studies (e.g., \cite{dong2017a}) show the heightened collaborations among the US, EU and China in recent years, we also tease apart the outputs and impacts resulted from collaborations among these three regions. Generally speaking, all articles are first reduced down to the unique countries/regions of their author affiliations without weightings, namely, multiple affiliations from the same country/region are counted only once for the country/region. Each article associated with more than one country/region is viewed as a product of collaborative research with equal contributions from all its countries/regions. These articles are counted for research output and impact of their respective regions as well as in the trans-regional research collaboration. For example, an article has two co-authors from the US, three from China, and another one from Europe. This article is counted once towards the regional output and impact of US, China, and Europe individually, and again for US-Europe, US-China, and Europe-China collaborations, respectively. The number of authors from each region has no effect in this study: this choice simplifies the considerations for authors affiliated with multiple affiliations of different regions.

\section{Results}
\label{sec:results}
The 2020 World Bank region codes for high-income countries (HIC’s), upper and lower middle-income countries (UMCs and LMCs), and low-income countries (LIC’s) are used to present the results. This study examines all STEM fields for all territories with World Bank region codes and the data are available for download in the Supplementary Material. 
\begin{figure}
    \centering
    \includegraphics[height=0.25\paperheight]{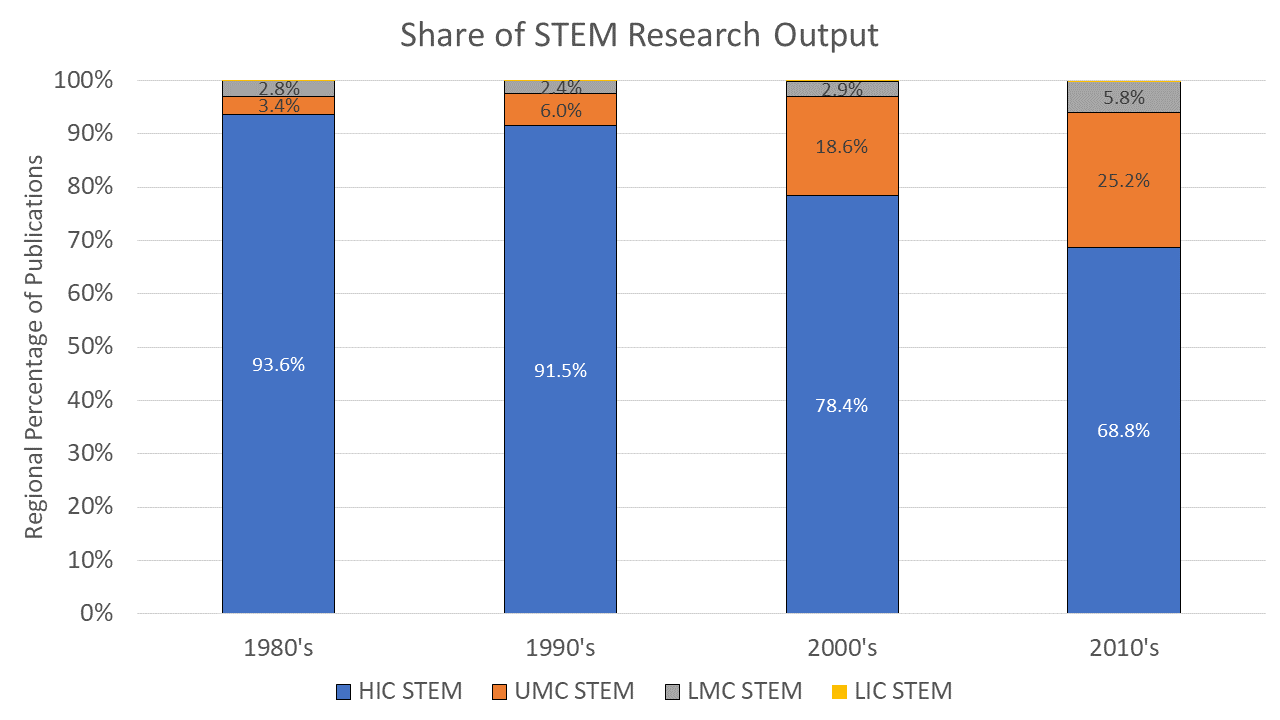}\newline
    \includegraphics[height=0.25\paperheight]{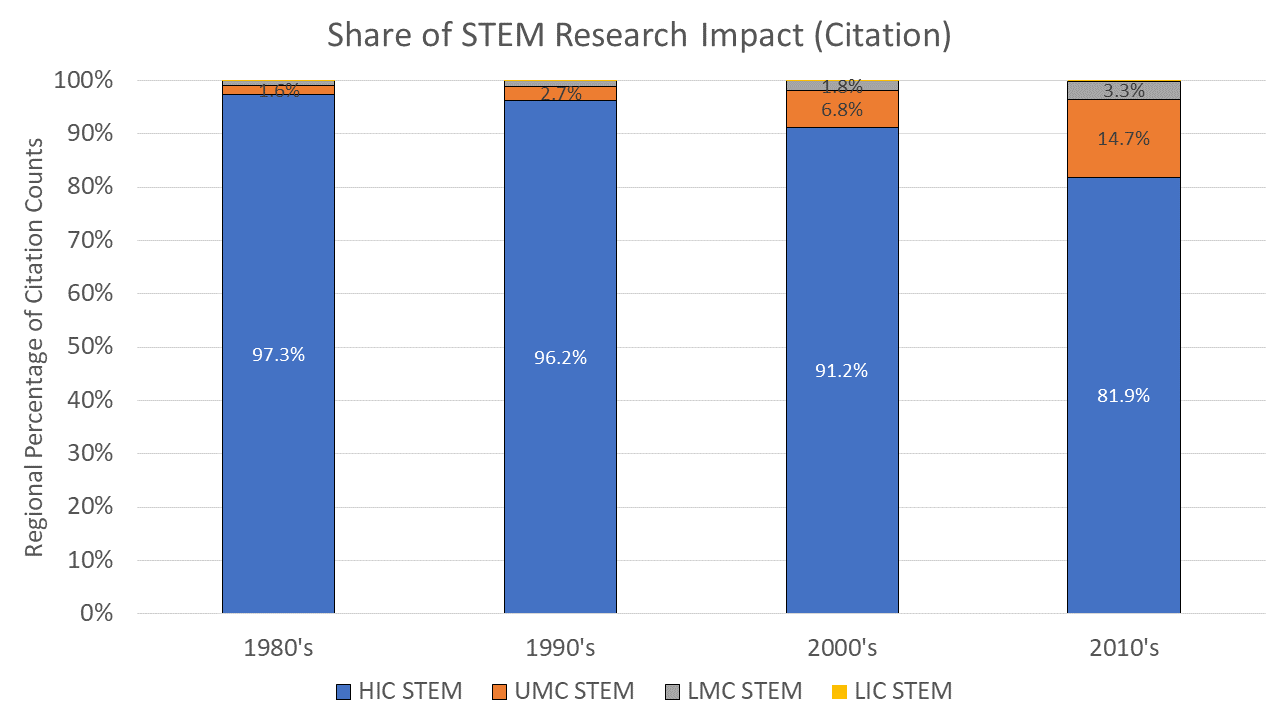}\newline
    \includegraphics[height=0.25\paperheight]{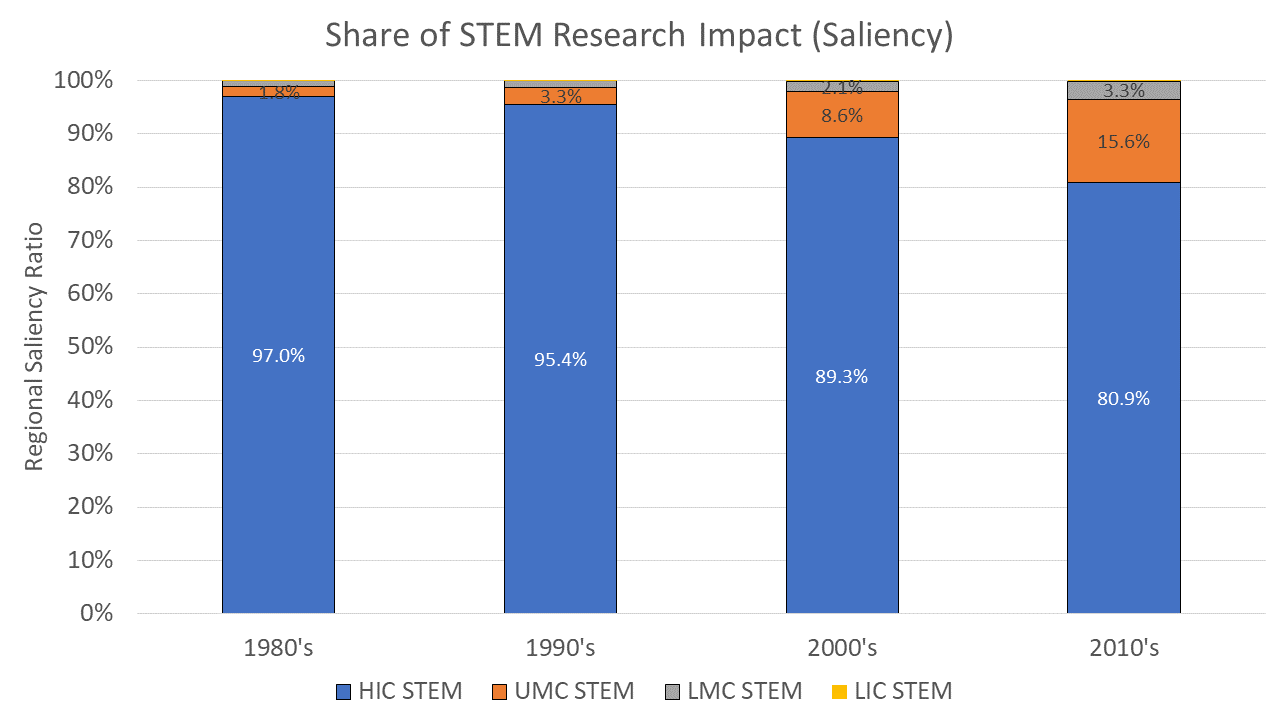}\newline
    \caption{The global STEM research output (top) and impact as measured in citation counts (middle) and saliencies (bottom) for the World Bank country groups for past 4 decades.}
    \label{fig:stemConvergence}
\end{figure}
Figure~\ref{fig:stemConvergence} shows the contribution margins in the research output and the impact, in either citation counts or saliencies, for these four groups over the past four decades.
At first glance, these results suggest a convergence in both research output and impact as the dominance of HICs is yielding to more contributions from the lower income countries. Particularly, the UMCs have accounted for more than one quarter of the world's research output and approximately 15\% of the impact, impressive growths from low single digit shares in the 1980's. Additionally, a necessary condition for convergence is lower income countries will grow much faster starting from a lower base. This is true between HICs and UMCs but does not hold for LMCs and LICs, suggesting a chasm may be forming between the UMCs and the LMCs. The latter can be clearly seen in Figure~\ref{fig:stemConvergence}: while LMCs and UMCs account for similar output and impact in 1980's, UMCs have produces almost 5 times more output and impact by 2010's than LMCs. Most importantly, the data show that highly aggregated results as presented in Figure~\ref{fig:stemConvergence} can be misleading for the following two reasons: first, the trends in STEM sub-fields exhibit considerable nuances and, second, the trends within distinct region groups also follow different patterns: while a convergence appears to take place within HICs, the opposite seems to be true within UMCs. To better illustrate these points, we pick a ``bellwether" country in each group, the United States (US) for the HICs and China (CN) for the UMCs, and examine their relative dominance in the groups they belong. 

\subsection{STEM Overall}\label{sec:stem}
\begin{figure}
    \centering
    \includegraphics[height=0.248\paperheight]{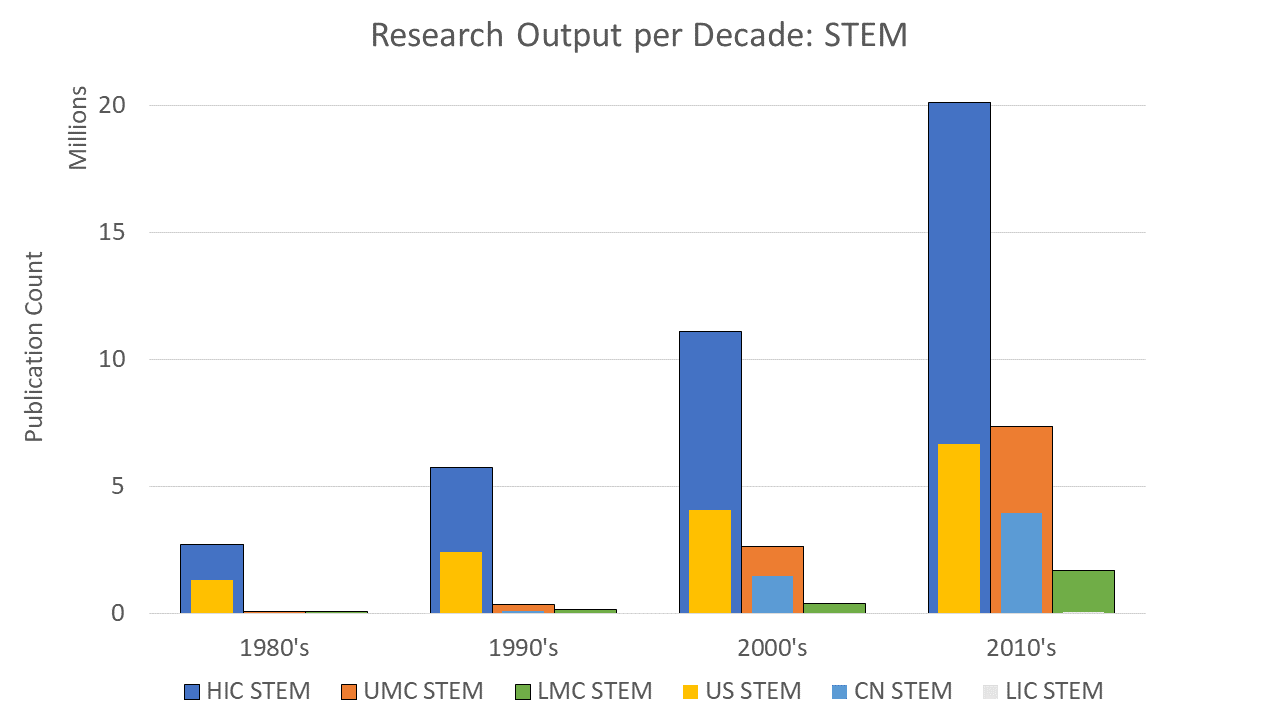}\newline
    \includegraphics[height=0.248\paperheight]{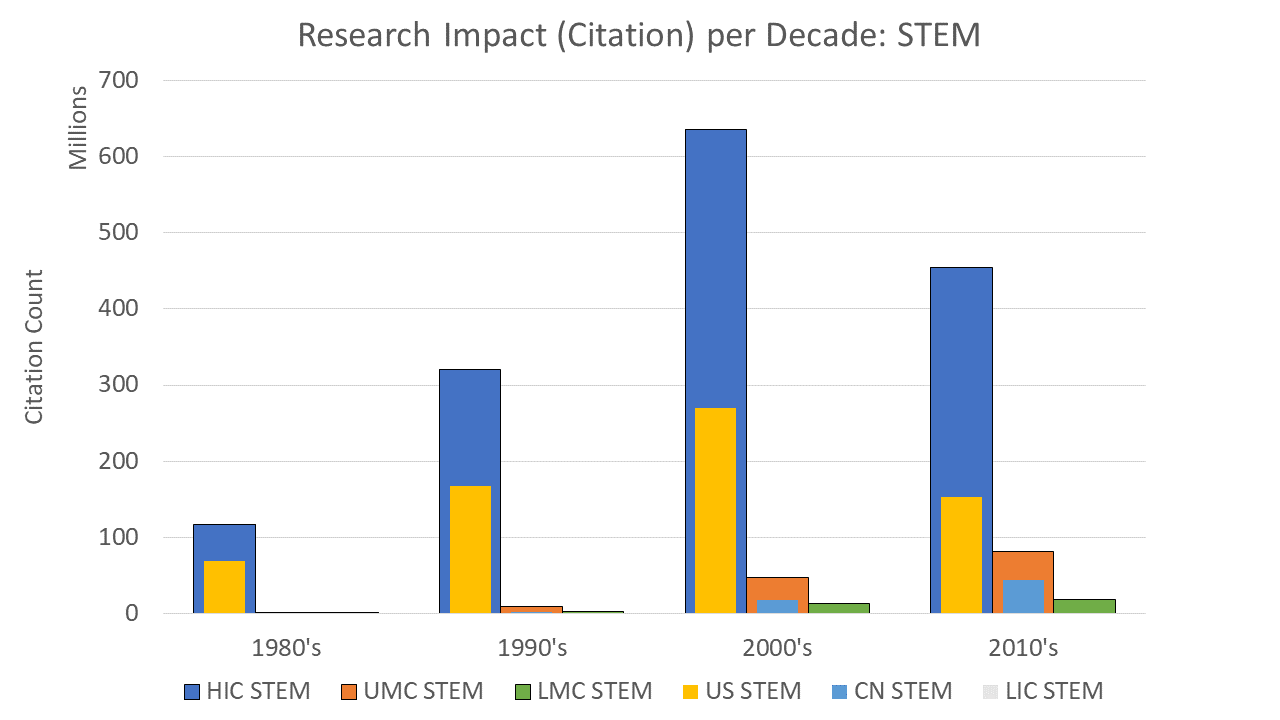}\newline
    \includegraphics[height=0.248\paperheight]{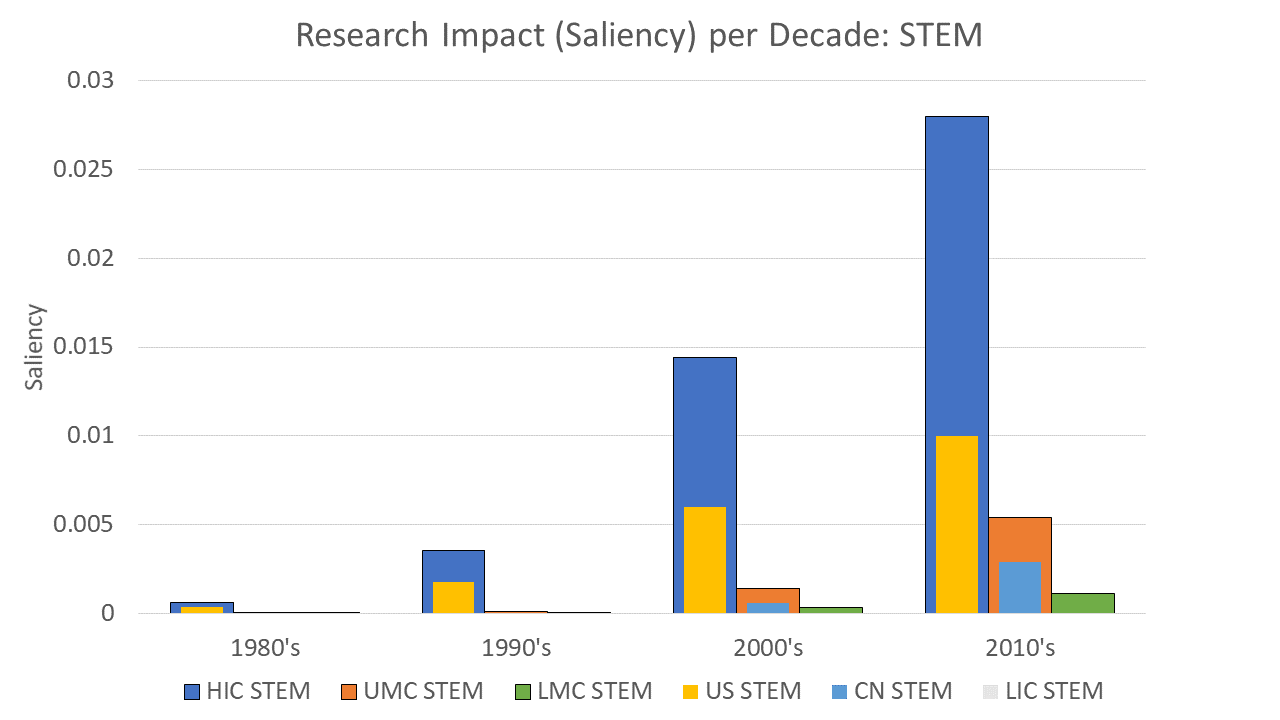}\newline
    \caption{STEM research output, as measured by publication count (top), and impact, measured by citation counts (middle) and saliency (bottom), for country groups per World Bank regional codes for the past four decades.}
    \label{fig:stem}
\end{figure}

Figure~\ref{fig:stem} shows the overall STEM research output and impact for all country groups for the four decades straddling the turn of the century. As can be seen in Figure~\ref{fig:stem}, groups ranging from HICs through LICs have all produced more scholarly publications growing at exponential rates, with lower income groups seeing a higher growth rate off a lower base. For example, the three decade-over-decade growth rates for HICs and LICs are 112.6\%, 92.5\%, 81.2\% and 170.1\%, 187.7\%, 281.8\%, respectively. While these growth rates may appear to meet the theoretical requirement for a convergence, the growth rates of LICs, nevertheless, are eclipsed by the middle income countries. In fact, most amazingly is the output from UMCs that, in 1980's, was more than an order of the magnitude behind HICs but has dramatically narrowed the gap in the most recent decade. One might take this development as an encouraging sign of convergence, yet the data also show the catching up of UMCs is at the price of a widening gap between UMCs and LMCs, from virtually non-existent in 1980's to more than half of an order of the magnitude in the most recent decade. Consequently, the output gaps between higher and lower incoming regions show little sign of narrowing, suggesting the catch-up effect has yet to emerge a the global scale. 

Notably, the convergence within individual groups, which can be seen by comparing the contribution ratios from the bellwether countries, depicts diametrical trends between the two upper income groups. Within HICs, the ratio of US output has reduced, suggesting the growing contributions from European and other high income countries have lessened the dominance of US. For UMCs, on the other hand, the opposite is true as China's research output is accounting for an ever higher ratio of the group's output (reaching more than 50\% in the recent two decades), making China \emph{the} undisputed dominant country among the UMCs. In fact, China alone has even surpassed the combined output of all LMCs since the first decade of the $21^{th}$ century.

A second and related question is whether there is a convergence in research impact. Data points that may shed lights on the answer are shown in Figure~\ref{fig:stem} where the conventional citation count and the novel saliency measures are used to quantify the community-wide recognition of impact (Sec.~\ref{sec:data}). Here, the citation count measure makes it difficult to arrive at a definitive conclusion as the citation counts of HICs, trending downwards, are confounded by the temporal bias against newer reports and cannot support the argument that the recent research from HICs has lower impact. It is, however, remarkable to note that the citation measure for UMCs bucks the trend and continues to show growth despite of the same temporal bias. 

In contrast, after adjusting for the temporal bias, the saliency measure shows that all income groups have produced research with increasing impact and the HICs, although with a higher base, continue to dominate the growth. Notably, the gap between HICs and UMCs has been narrowing. The impact from the United States alone, for example, still doubles that of all of the combined work from the UMCs even though its share is lessened considerably among its HICs peers. The results suggest a convergence in research impact among the HICs. However, like the case of research output, the trend in UMCs shows the increasingly dominant role of its bellwether country, China, that in the most recent decade has seen its saliency ratio grow to over 50\%, i.e., accounting for more than half of the research impact from the UMC group. Again, the research from lower income countries of LMCs and LICs has yet to play a significant role in the global research arena. The data support the argument that the lower income countries lag even more in their impact in recent decades, with China alone surpassing all their combined work by a wide margin.
. 
\subsection{Tales of three sub-fields} \label{sec:subfields}
The highly aggregated results of STEM run the risk of painting the picture with too broad a brush as the data show uneven rates of convergence in the STEM sub-fields: While there are fields seeing more pronounced convergence as in AI, Chemistry and Materials Science at least among the higher income categories, there are still many fields, e.g., Medicine, where the historical leaders remain overly dominant. To illustrate this point, we selectively discuss the results of three representative fields, AI, Medicine and Materials Science, respectively, noting that the data in the supplementary material can provide deeper evidence for other fields.

\subsubsection{Artificial Intelligence (AI)} \label{sec:ai}
\begin{figure}
    \centering
    \includegraphics[height=0.245\paperheight]{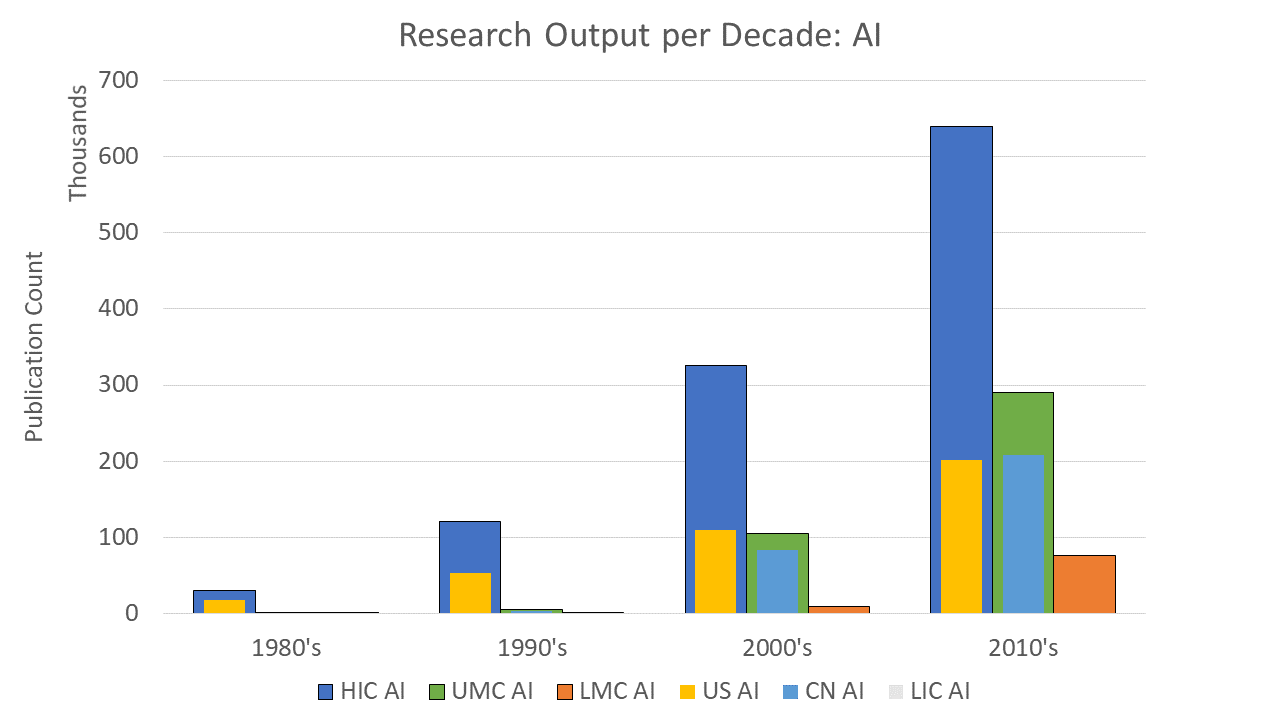}\newline
    \includegraphics[height=0.245\paperheight]{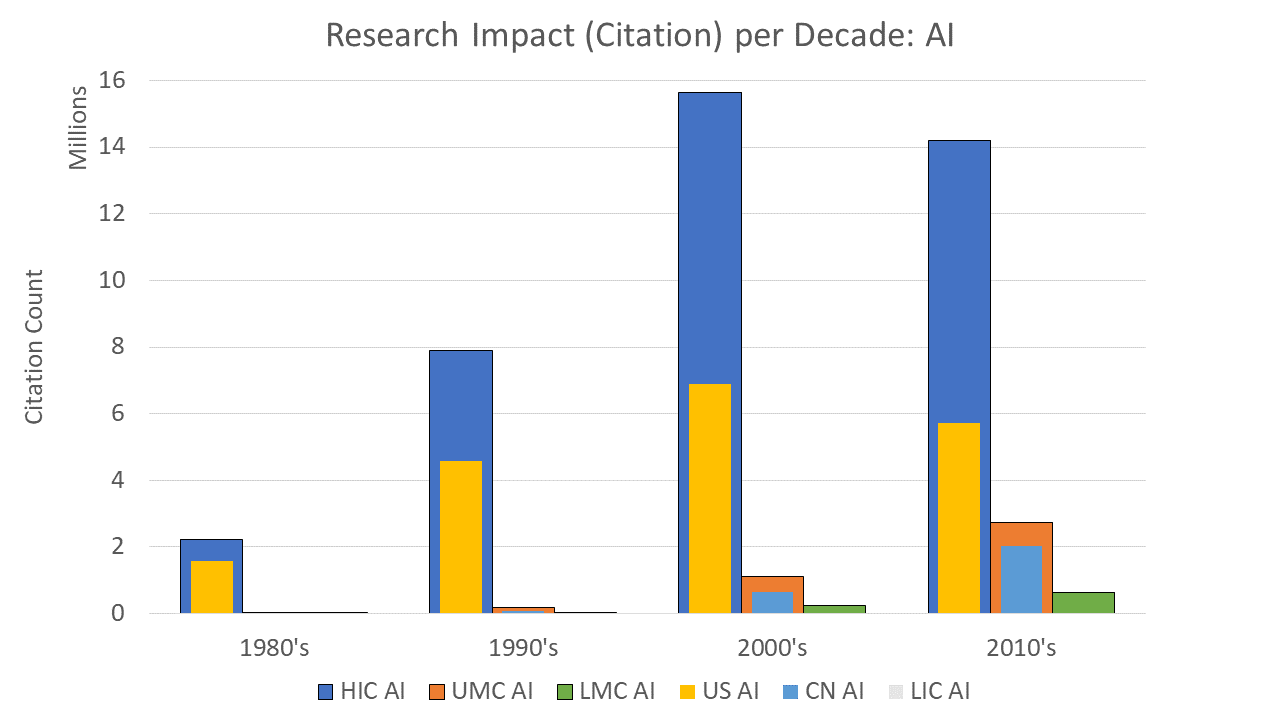}\newline
    \includegraphics[height=0.245\paperheight]{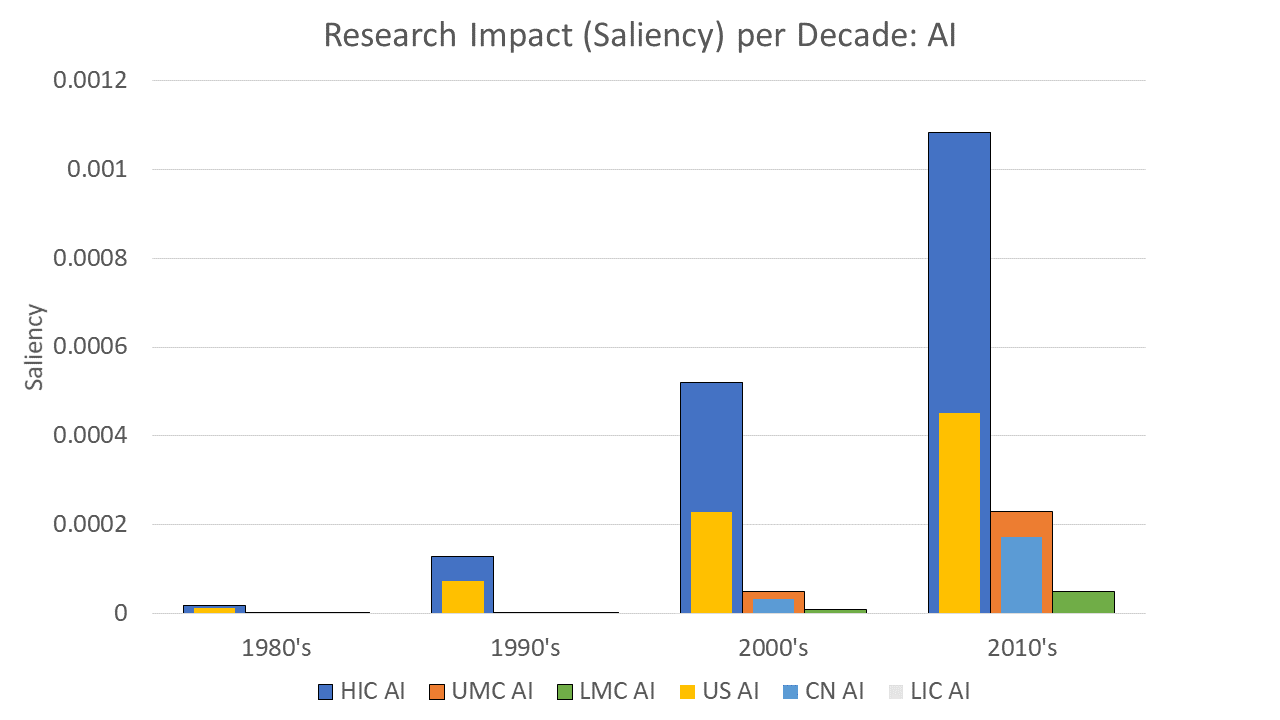}\newline
    \caption{Research output, measured by publication count (top), and impact, by citation count (middle) and saliency (bottom), for the field of AI that is one of the fields where the globalization convergence in research output seems to accelerate within HICs and UMCs in recent decades, yet the gain does not translate proportionally to research impact.}
    \label{fig:ai}
\end{figure}
AI, a field rooted in Computer Science but with rich history of interdisciplinary cross-pollination, is an example in which the convergence in research output outpaces STEM, and the sign of a convergence in research impact is modest but encouraging. As can be seen in Figure~\ref{fig:ai}, the current landscape bears little resemblance to that of the last century where the field was largely propelled by the HICs and, within it, dominated by the US in research impact but not in output. During the four decades, the publication and saliency ratios of HICs change from 93.6\% and 97.0\% in 1980's to 68.8\% and 80.9\%, respectively.
Within HICs, the quality of AI research has caught up since the turn of the century that the US has accounted for less then half of the impact. To be more specific, the US accounts for 56.9\% of research output and has a saliency ratio 70.8\% in 1980's, but in the most recent decade these two indicators have lowered to 20.0\% and 33.2\%, respectively.

In the meantime, the lower income countries have produced more research in AI, especially the impressive gain by the UMC group. UMCs have collectively generated research output larger than the US alone and, by the second decade, the bellwether country in the group, China, has matched the US in research output. In terms of research impact, however, UMCs have made impressive progress but still have a significant room left in order to catch up with HICs. For the most recent decade, the impact of the US alone is approximately twice as high as the collective impact from UMCs. As in the case of general STEM, UMCs show scant signs of convergence as the group's research output and impact are largely dominated by China. The publication and saliency ratio of China have grown from 1.0\% and 0.7\% in 1980's to 20.6\% and 12.6\% in 2010's, respectively.

\subsubsection{Medicine} \label{sec:medicine}
\begin{figure}
    \centering
    \includegraphics[height=0.25\paperheight]{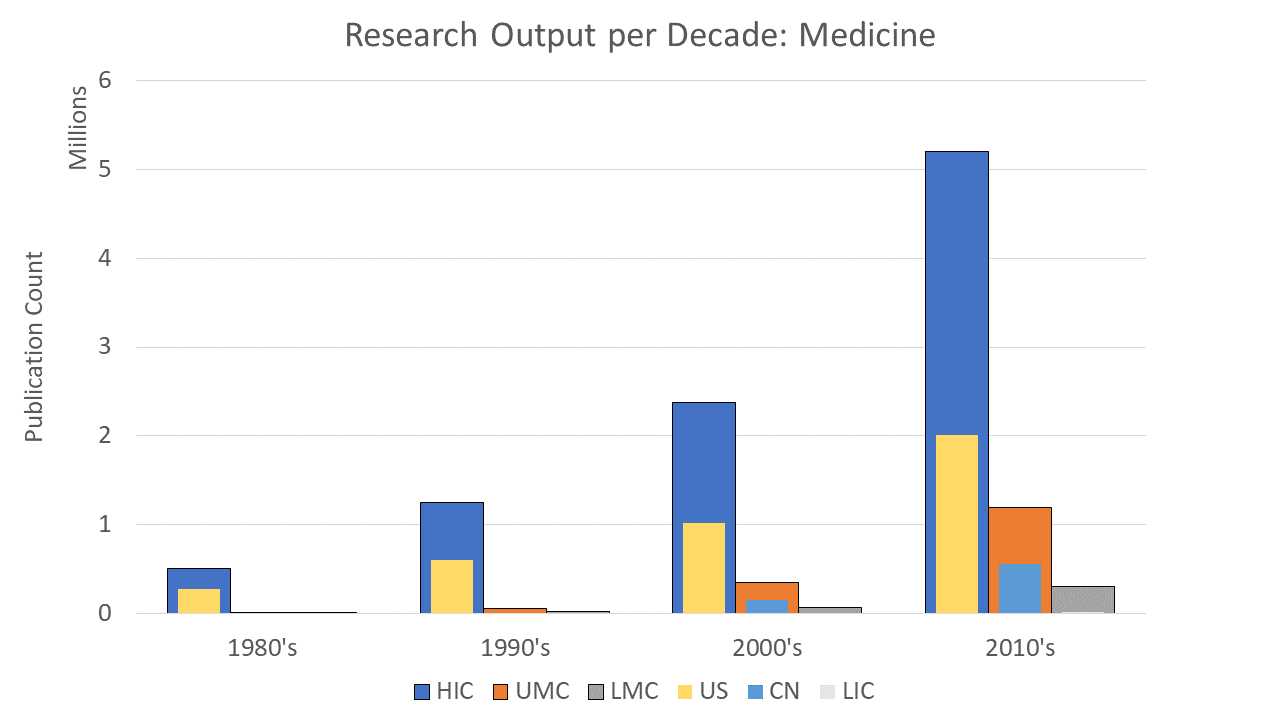}\newline
    \includegraphics[height=0.25\paperheight]{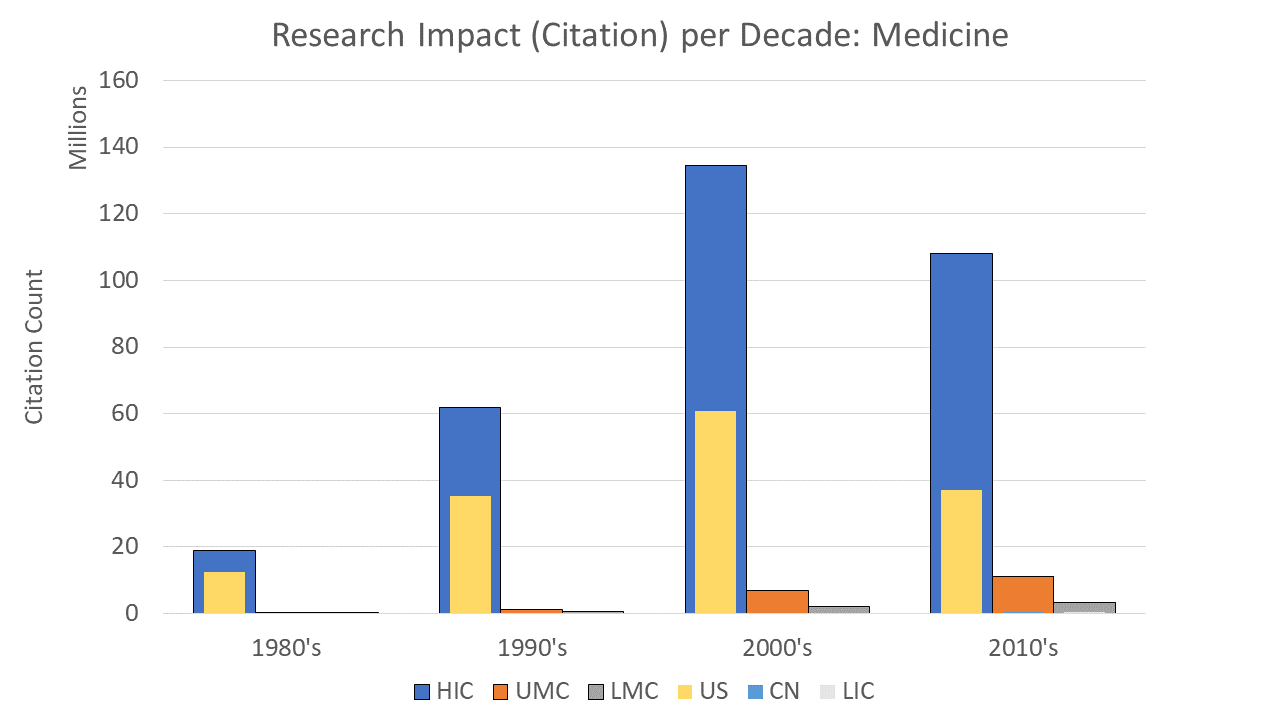}\newline
    \includegraphics[height=0.25\paperheight]{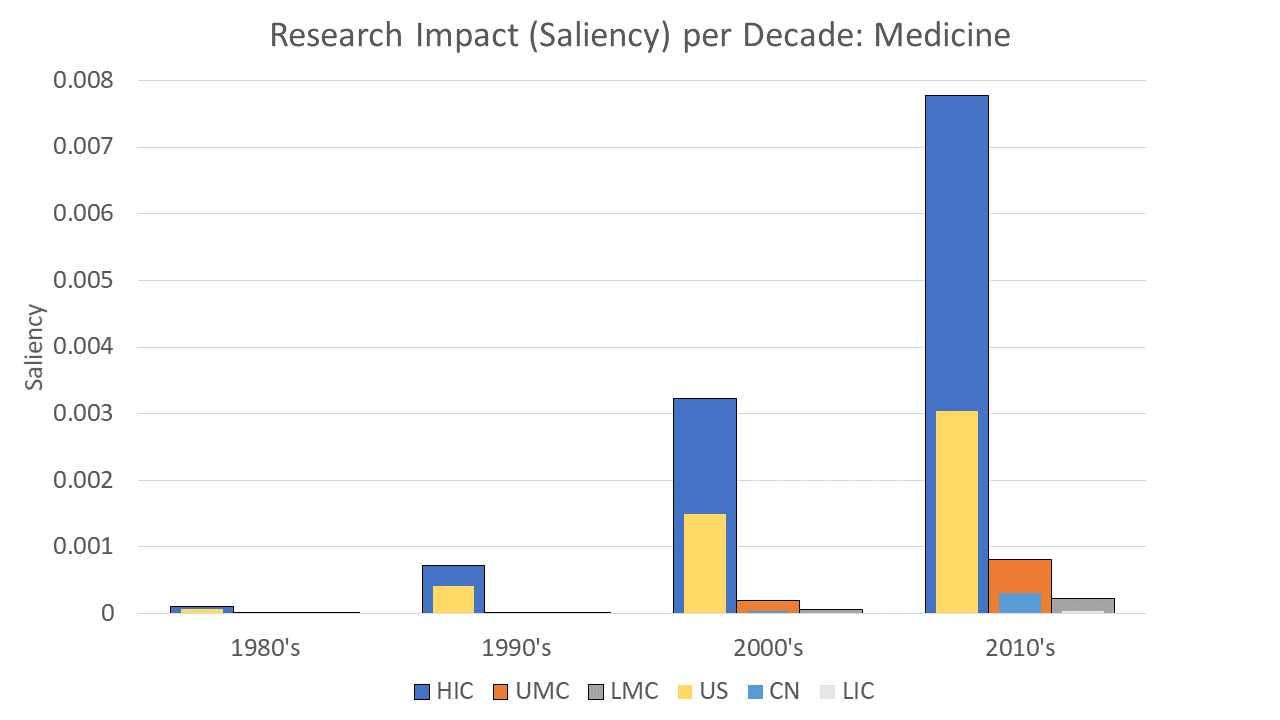}\newline
    \caption{Research output and impact for the field of Medicine, one that no significant convergence can be observed in recent decades, either in research output or impact.}
    \label{fig:medicine}
\end{figure}
Medicine is one of the fields where the globalization seems to have changed little the origins of research output and impact. As shown in Figure~\ref{fig:medicine}, the field remains dominated by the HICs, even though the lower incoming countries have collectively made a substantial gains. Within the four decades, HICs' research output has changed from 95.8\% to 77.3\%, and research impact from 97.7\% to 87.9\%, with the US alone still far exceeding the collective impact from the lower incoming countries.
\subsubsection{Materials Science} \label{sec:material}
\begin{figure}
    \centering
    \includegraphics[height=0.25\paperheight]{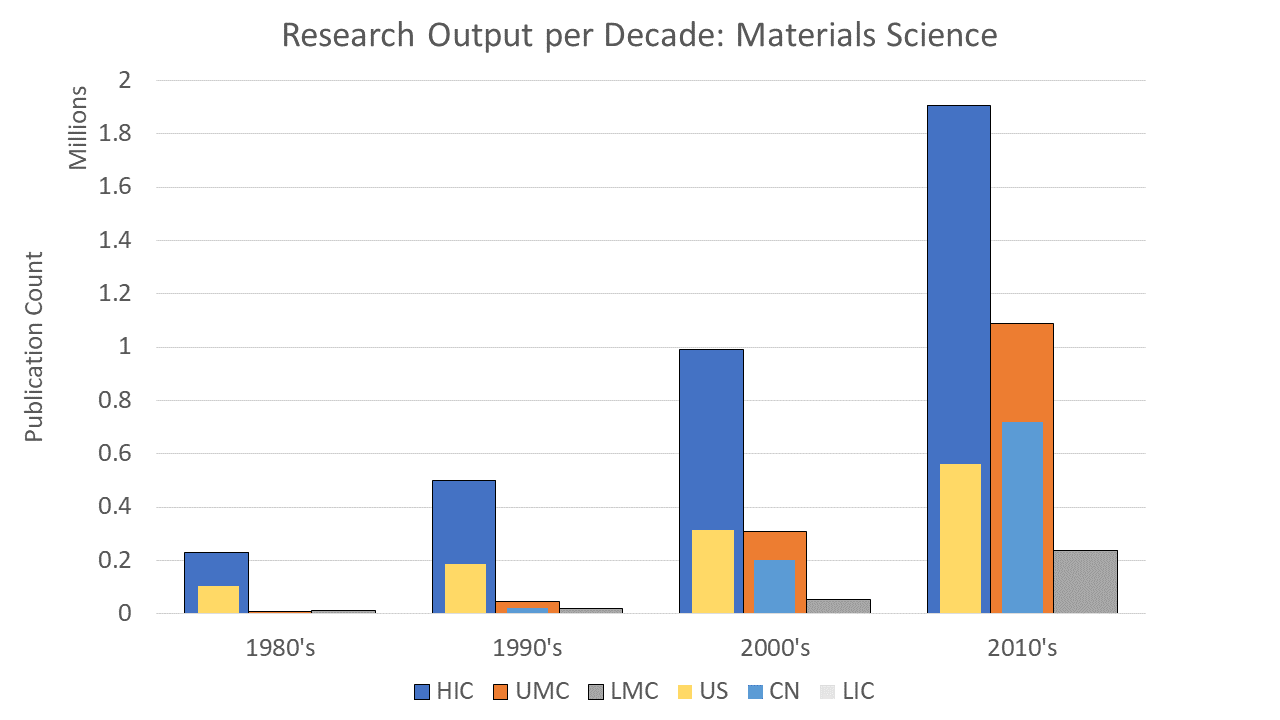}\newline
    \includegraphics[height=0.25\paperheight]{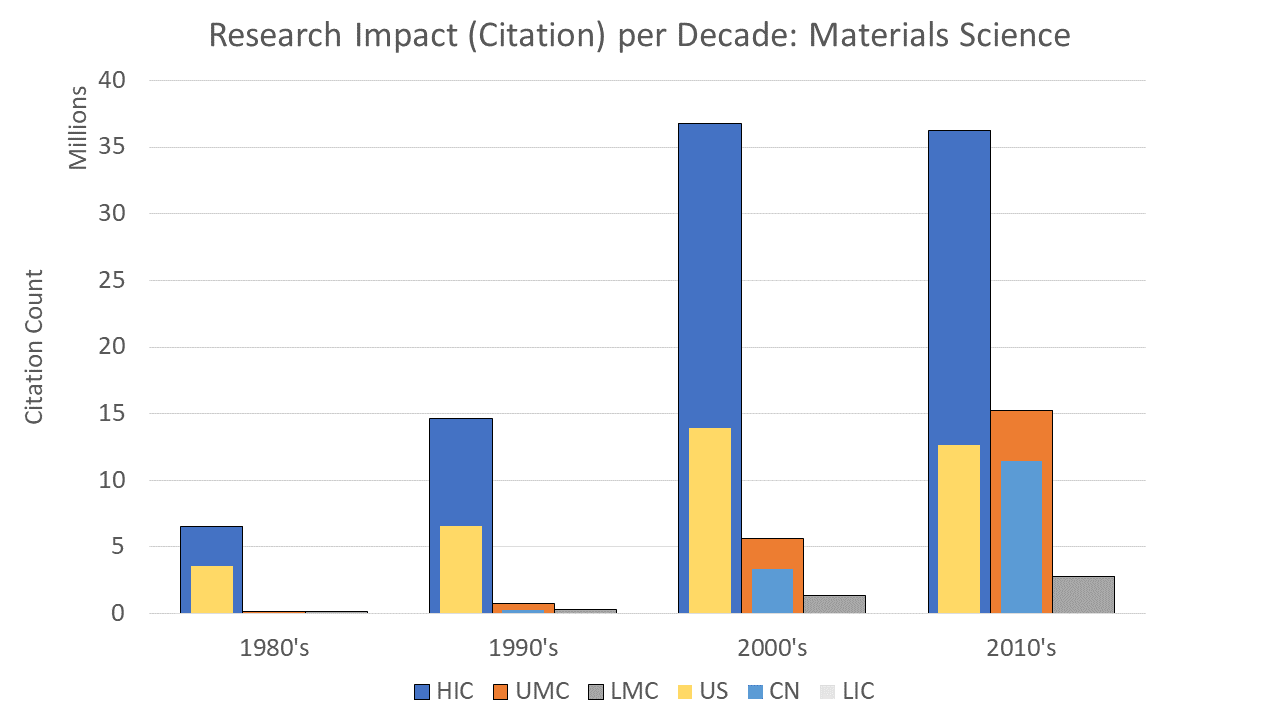}\newline
    \includegraphics[height=0.25\paperheight]{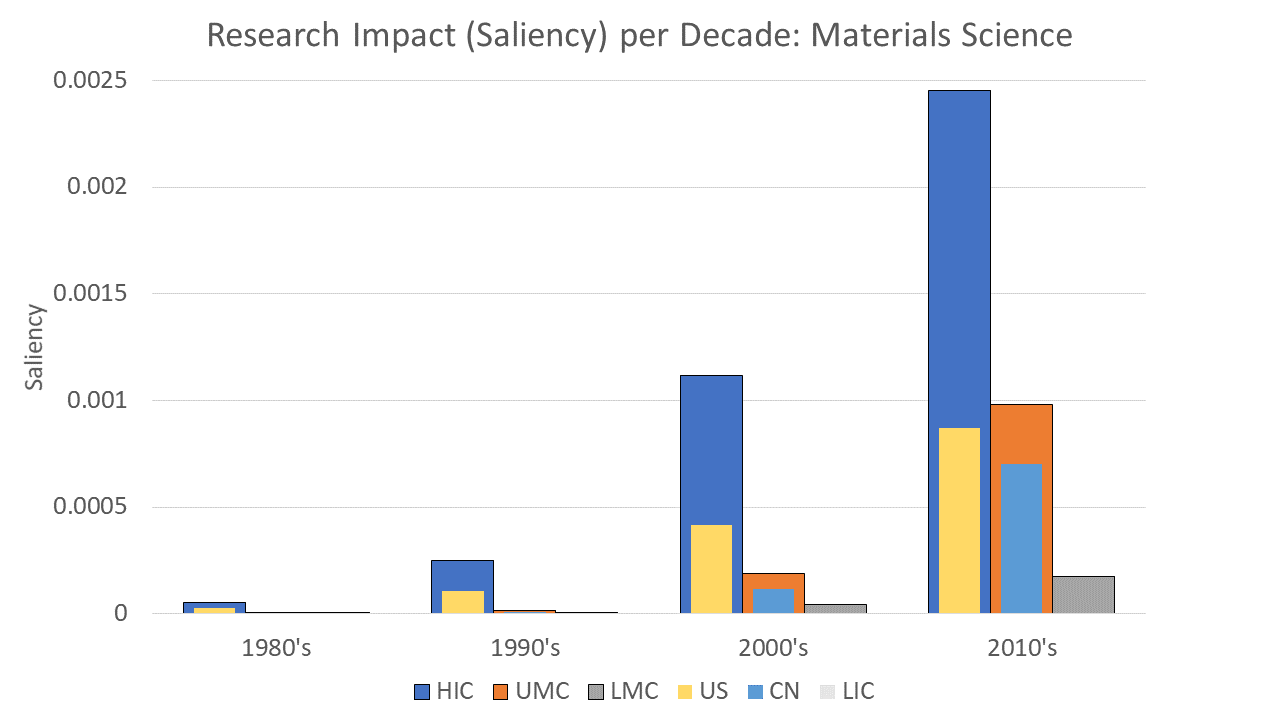}\newline
    \caption{Research output and impact for the field of Materials Science, where the UMCs have significantly narrowed the gaps with HICs in both output and impact.}
    \label{fig:material}
\end{figure}
Materials Science is one of the few fields where the sign of convergence is most observable in that no single country in the HICs has produced more research impact than the lower incoming countries combined. As can be seen in Figure~\ref{fig:material}, the field has seen the output and impact gains from the lower income countries that the publication and saliency ratios from the HICs are reduced from 91.9\% and 96.0\% in 1980's to 58.9\% and 67.9\%, respectively. Most remarkably, while the US alone still produced more output and impact than the combined work of the lower income countries in the first decade of the century, it is no longer the case in the most recent decade. In fact, China has overtaken US in output and is quickly catching up on impact. Notably, while the convergence within the HIC group is clear, the data suggest the same may also be true within the UMCs as the dominance of China seems to be loosened somewhat between the first and the second decade of the century. This observation, however, has to be taken with a grain of salt because it is based on a single data point, and there are limitations in our data that only consider publications addressed to the international audience, i.e., written in English. The latter point is further elaborated in Sec.~\ref{sec:discussion}.

\subsection{The role of cross-border collaboration} \label{sec:collaborations}
\begin{figure}
    \centering
    \includegraphics[width=0.49\textwidth]{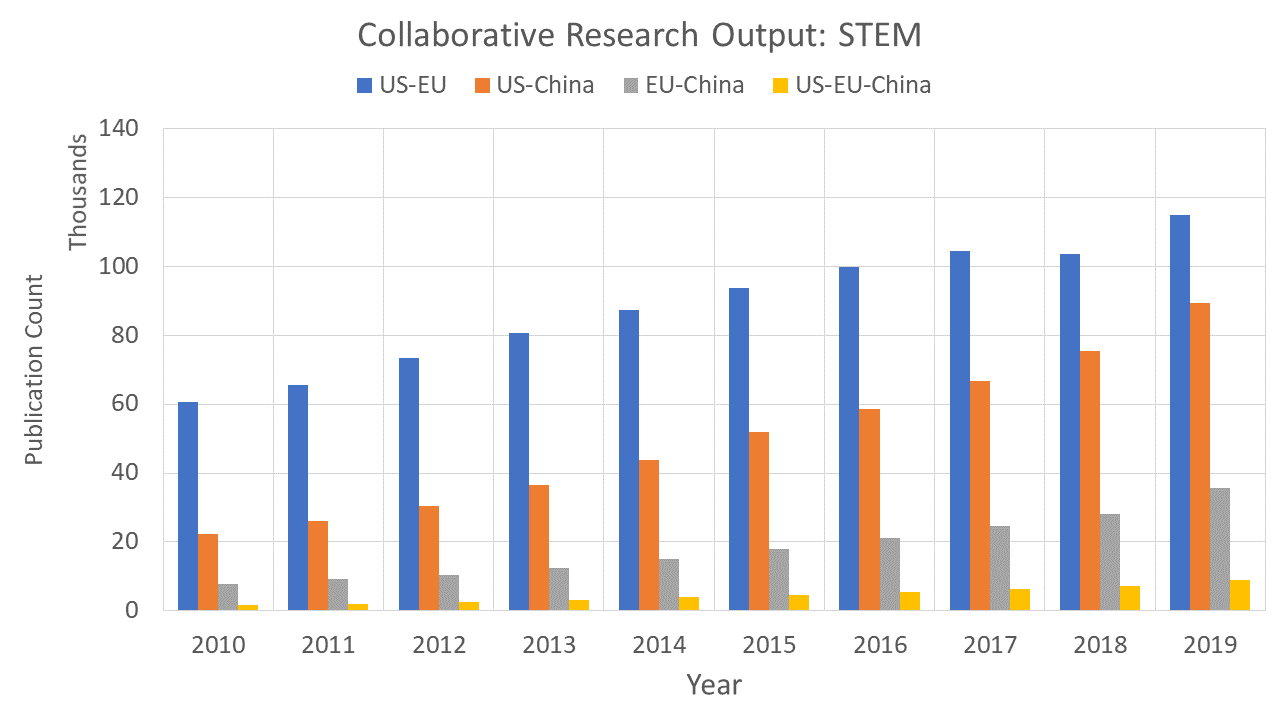}
    \includegraphics[width=0.49\textwidth]{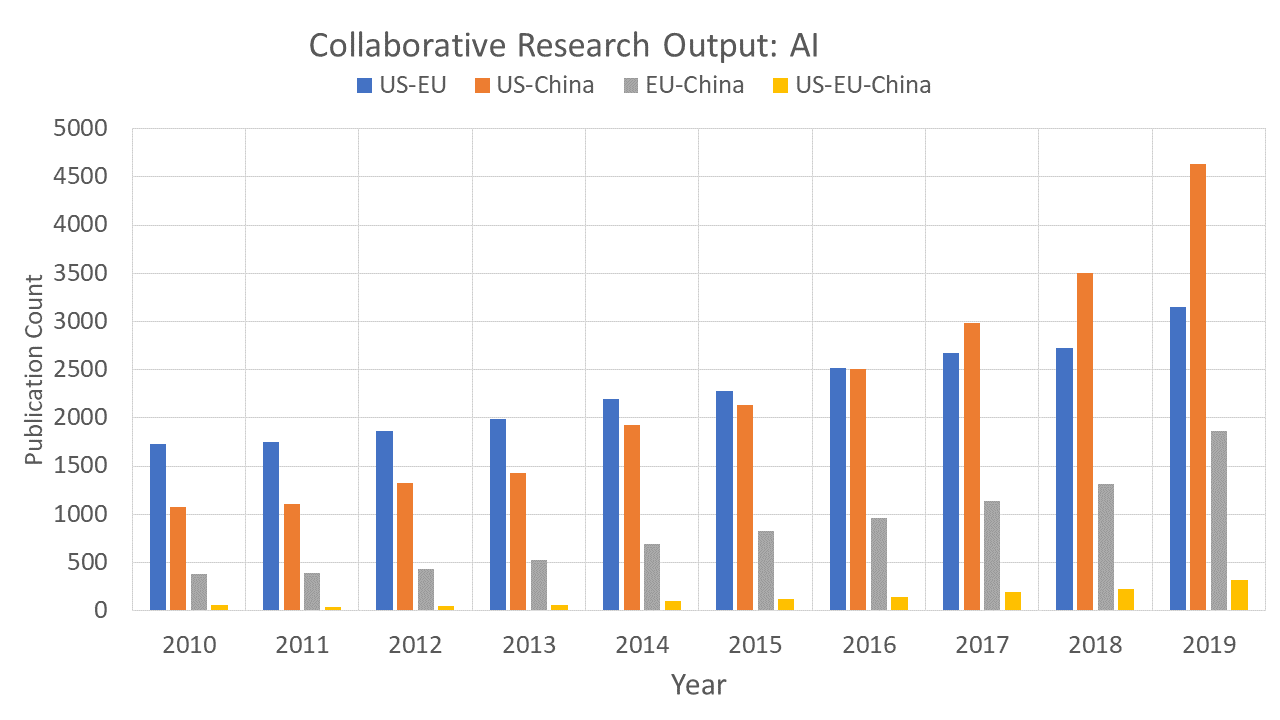}\newline
    \includegraphics[width=0.49\textwidth]{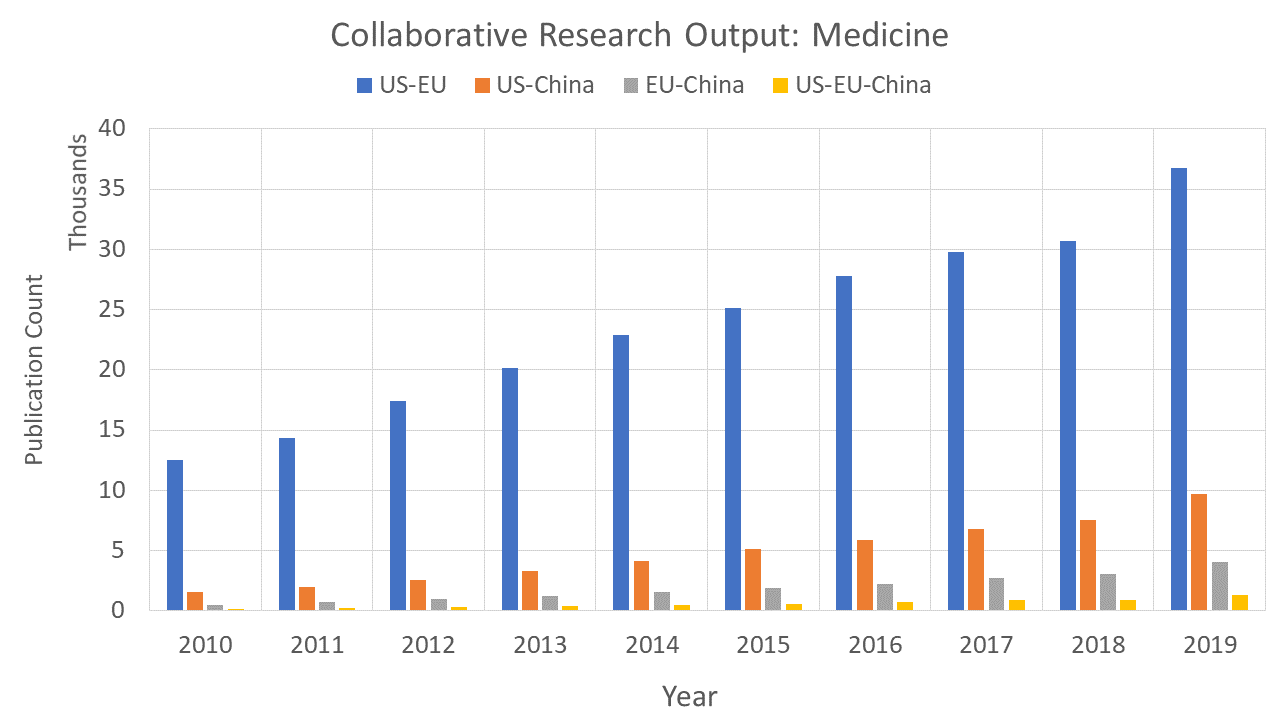}
    \includegraphics[width=0.49\textwidth]{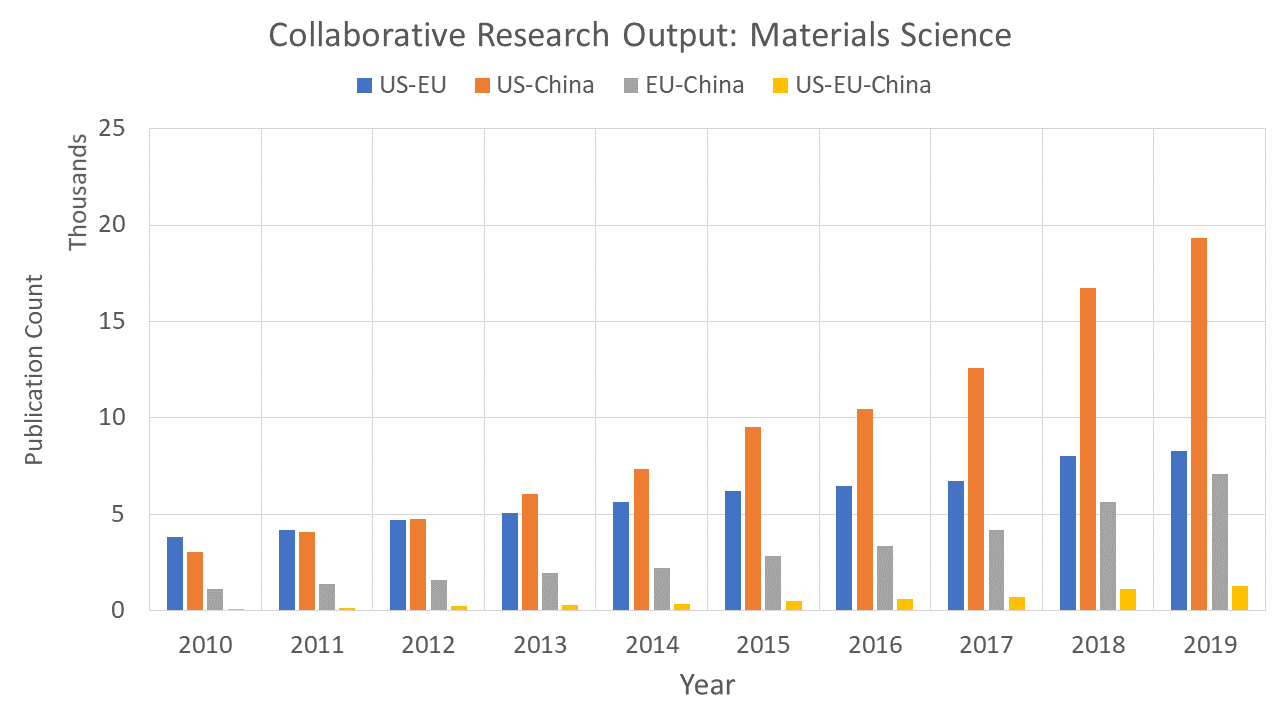}\newline
    \caption{Research output of global collaborations for the most recent decade for STEM research and three representative sub-fields. The output is measured by the number of research articles published in the years from 2010 through 2019.}
    \label{fig:CollabOuput}
\end{figure}

At the root of globalization is collaboration across regional boundaries, and the momentum of such collaboration may be a leading indication of an eventual convergence. Literature abounds in reporting increasing research output as results of collaboration (Sec.~\ref{sec:intro}), and our data, as shown Figure~\ref{fig:CollabOuput}, confirm these observations. Aside from the quantity, we further examine the quality of collaborative research, measured by the saliency ratio of such work, and find the trends most encouraging. While the detailed data in the Supplementary Material offer in-depth insights into individual countries, the state of the global collaborations can be best described by contrasting the recent rise of China, be it the transpacific collaborations with the US or the Sino-European activities, against the traditional transatlantic research collaborations between the US and the European countries. Either in terms of output or impact, the trends in research collaborations bear remarkable parallels to the regional results described above.

Figure~\ref{fig:CollabOuput} shows the output as a result of transregional collaborations of the transatlantic (US-EU), the transpacific (US-CN), and the Sini-European research in the most recent decade. Generally speaking, STEM research has seen growing collaborations that track the growth of research output overall, and the transpacific collaborations seems to be gaining on the traditional transatlantic ones. However, the highly aggregated STEM results do not highlight nuanced details varying significantly among sub-fields. For instance, Medicine is the area where the pattern has hardly evolved from the dominant transatlantic collaborations dated back for decades, while the transpacific collaborations have noticeably overtaken transatlantic ones in AI and Materials Science. The sign of the catch-up effect in research output (at least between the HICs and the UMCs) is most and least observable in the Materials Science and Medicine, with AI in between.
\begin{figure}
    \centering
    \includegraphics[width=0.49\textwidth]{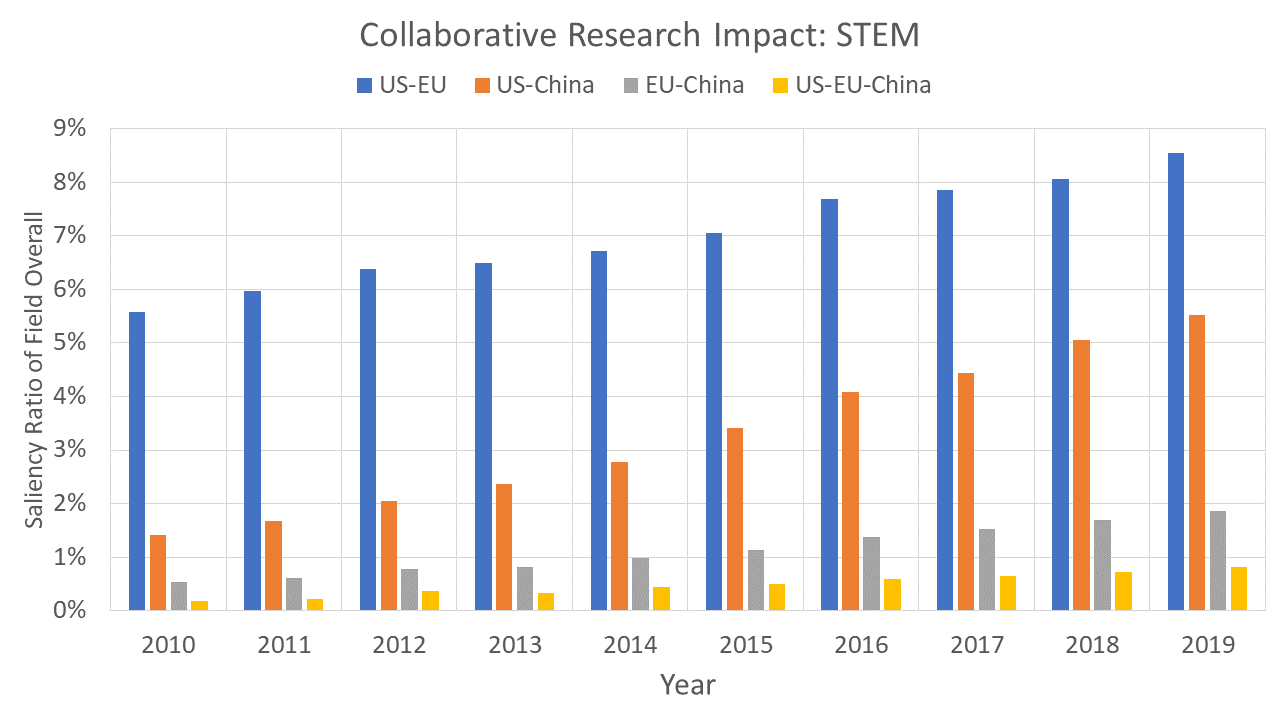} \includegraphics[width=0.49\textwidth]{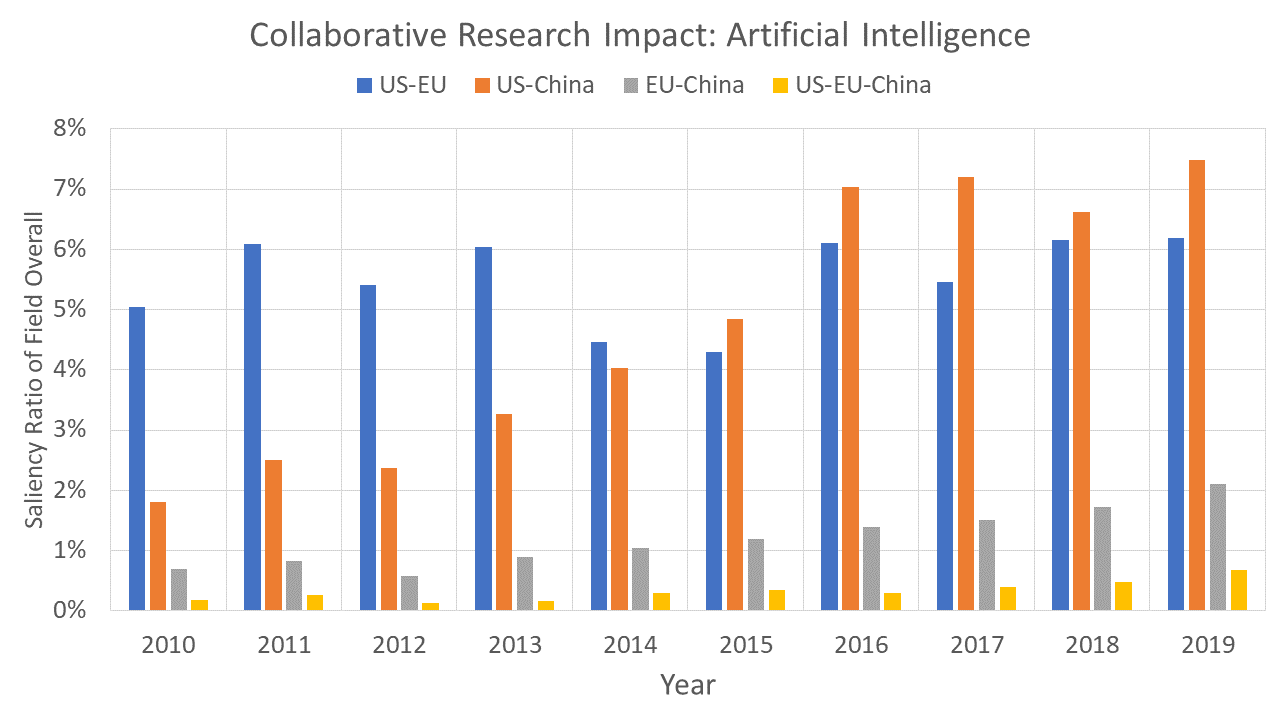}\newline
    \includegraphics[width=0.49\textwidth]{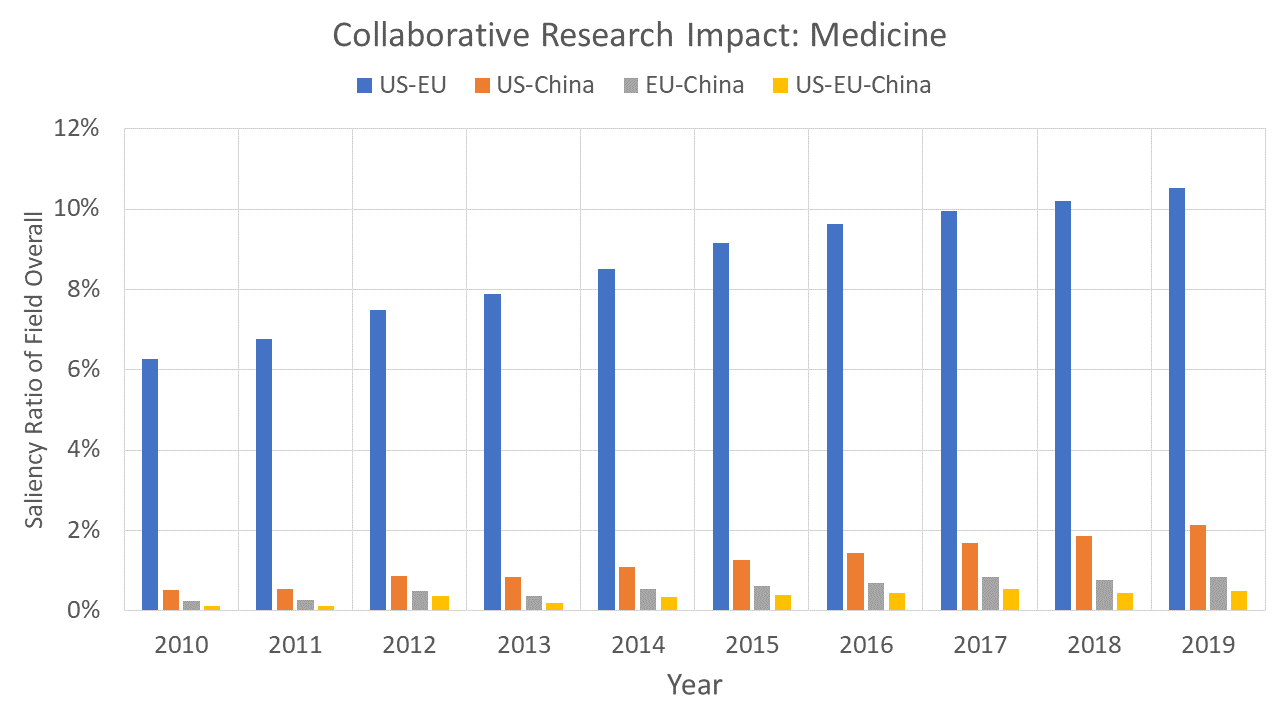} \includegraphics[width=0.49\textwidth]{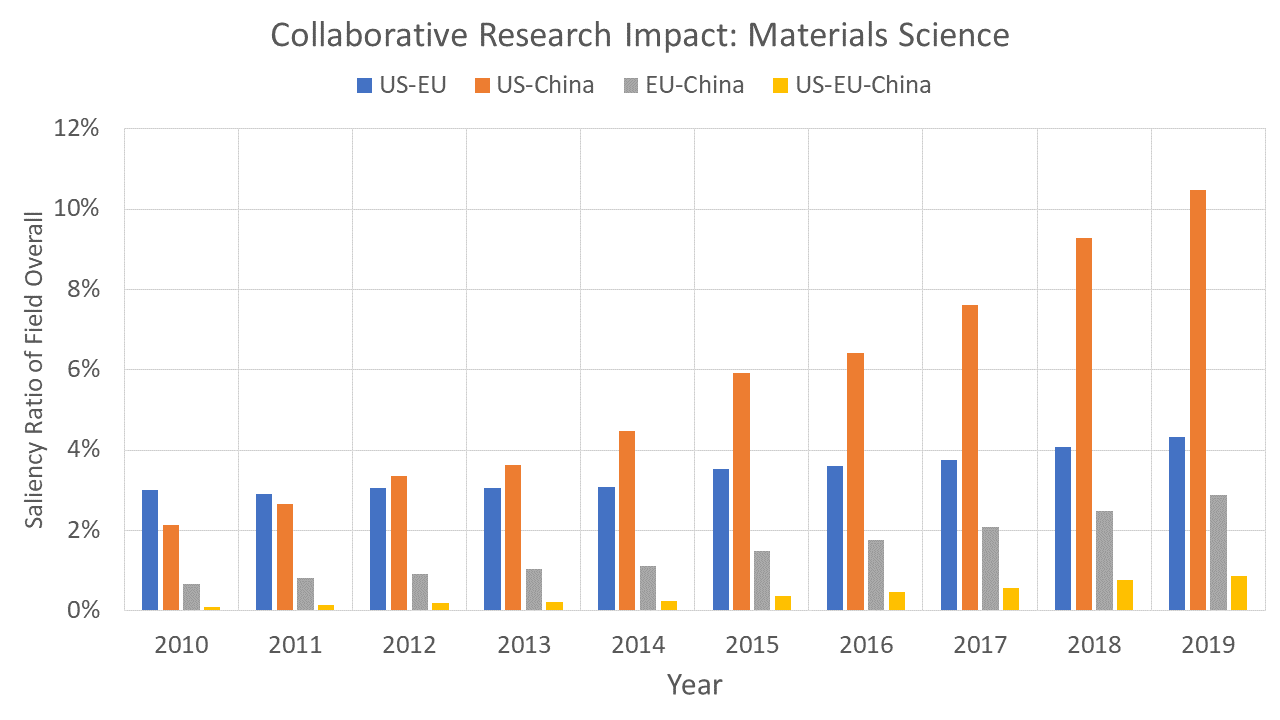}\newline
    \caption{Research impact, measured in saliency ratio of the STEM overall impact, for collaborative research for the past decade.}
    \label{fig:CollabImpact}
\end{figure}

Similarly in terms of impact, the data suggest that STEM research collaborations are playing an increasingly important role relative to the work produced in the same year. This can be seen in Figure~\ref{fig:CollabImpact} as the saliency of collaborative work is accounting more share of the impact of the same year. Indeed, there are fields such as Medicine where the transatlantic collaborations still dominate in impact, but the fields of AI and Materials Science have seen transpacific collaborations upend the traditional norms. The transpacific collaborations in Materials Science, the research impact from transpacific collaborations has become more than double that of the transatlantic ones, and the transpacific collaborations in AI have surpassed transatlantic collaborations in research impact since 2015. In fact, the data show most of the impact gains in China's AI research in recent decades can be largely attributed to research with global collaborations. The fact that collaborative work increasingly accounts for a higher saliency ratio suggest high impact research is likely a result of global collaboration, an encouraging indication to support a convergence of globalized research.

\begin{figure}
    \centering
    \includegraphics[width=0.49\textwidth]{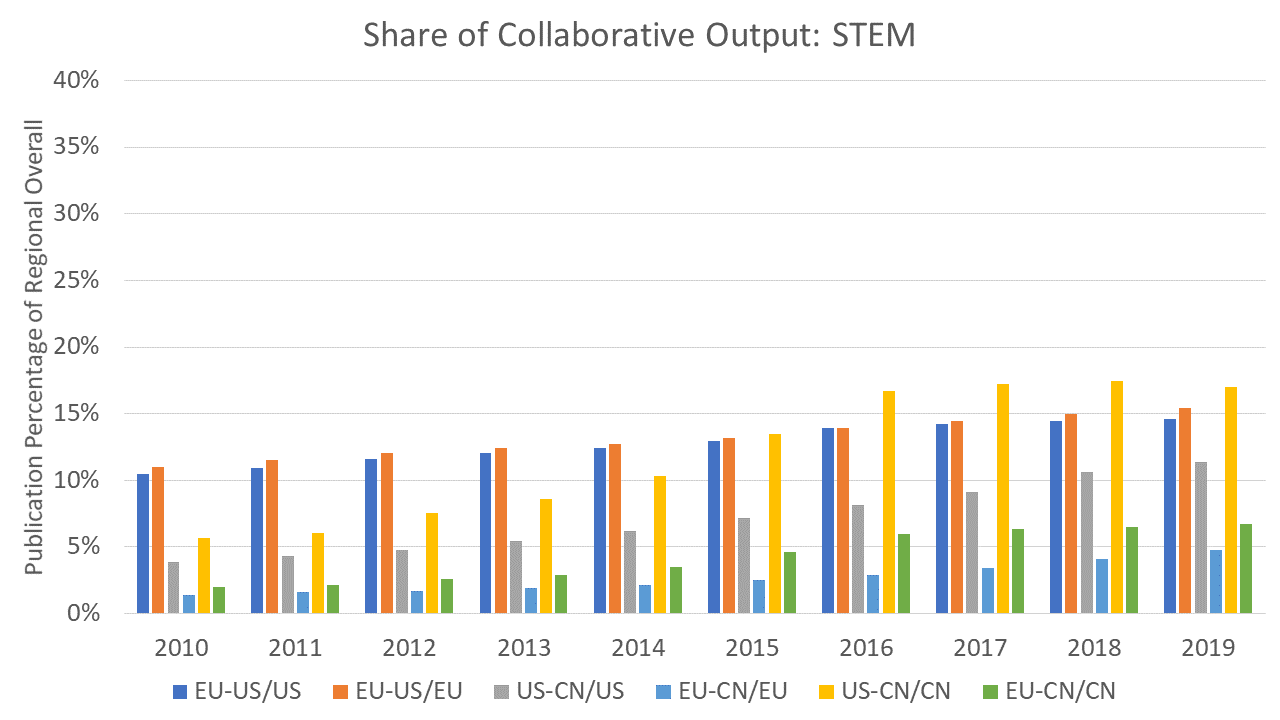} \includegraphics[width=0.49\textwidth]{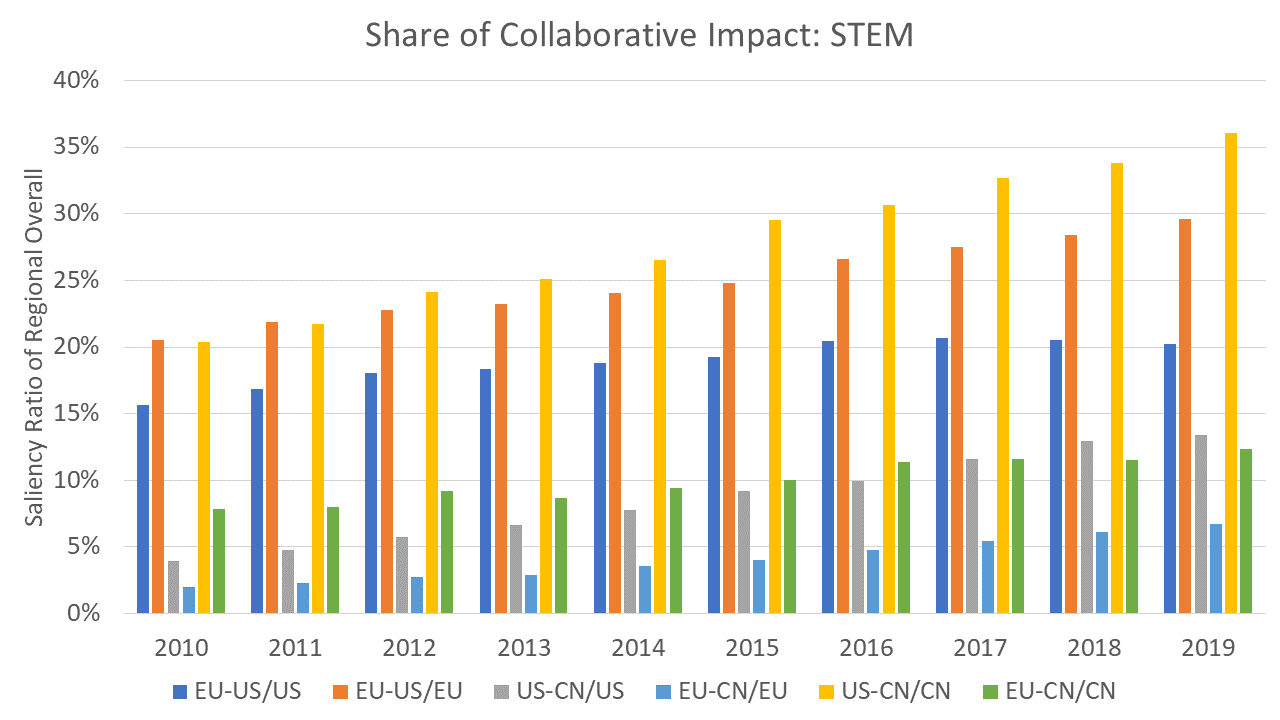}\newline
    \caption{Shares of STEM research output (left) and impact (right) as a result of transregional collaboration with respect to individual countries of US, EU, and China for the past decade. The share of research output and impact are measured by the percentage of published articles and the saliency ratio, respectively.}
    \label{fig:CountryCollabSTEM}
\end{figure}

\begin{figure}
    \centering
    \includegraphics[width=0.49\textwidth]{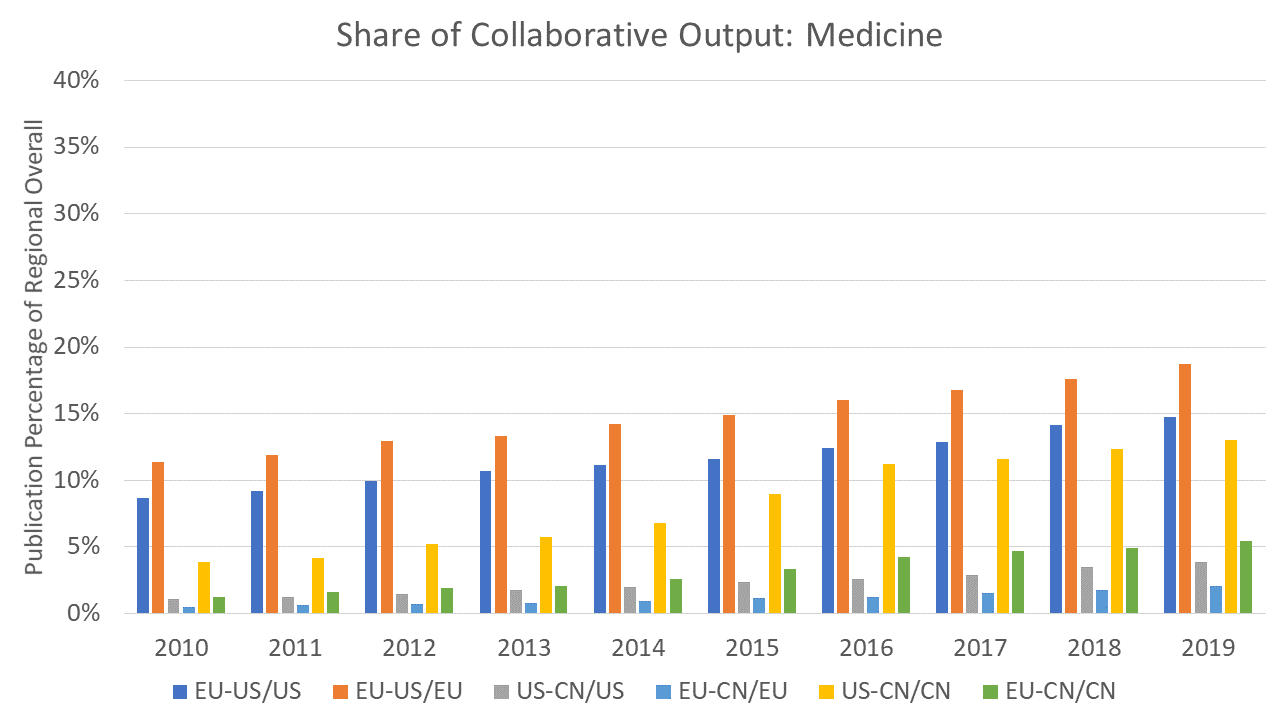} \includegraphics[width=0.49\textwidth]{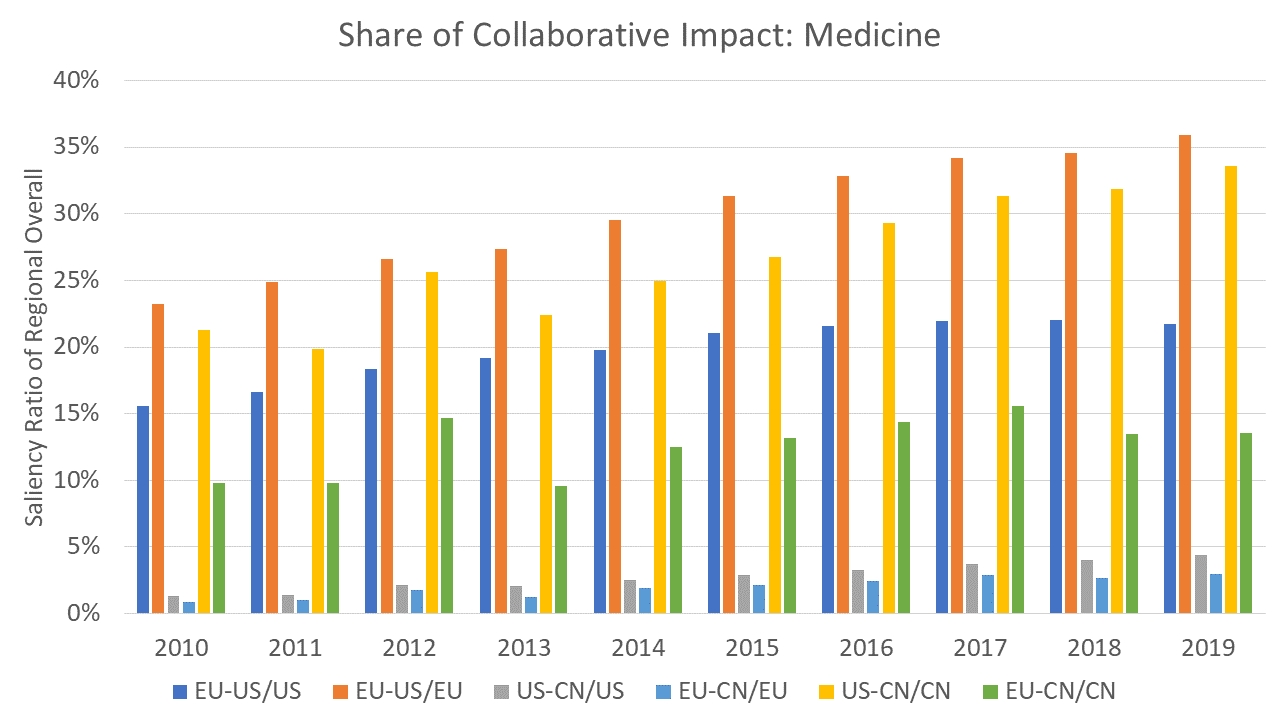}\newline
    \includegraphics[width=0.49\textwidth]{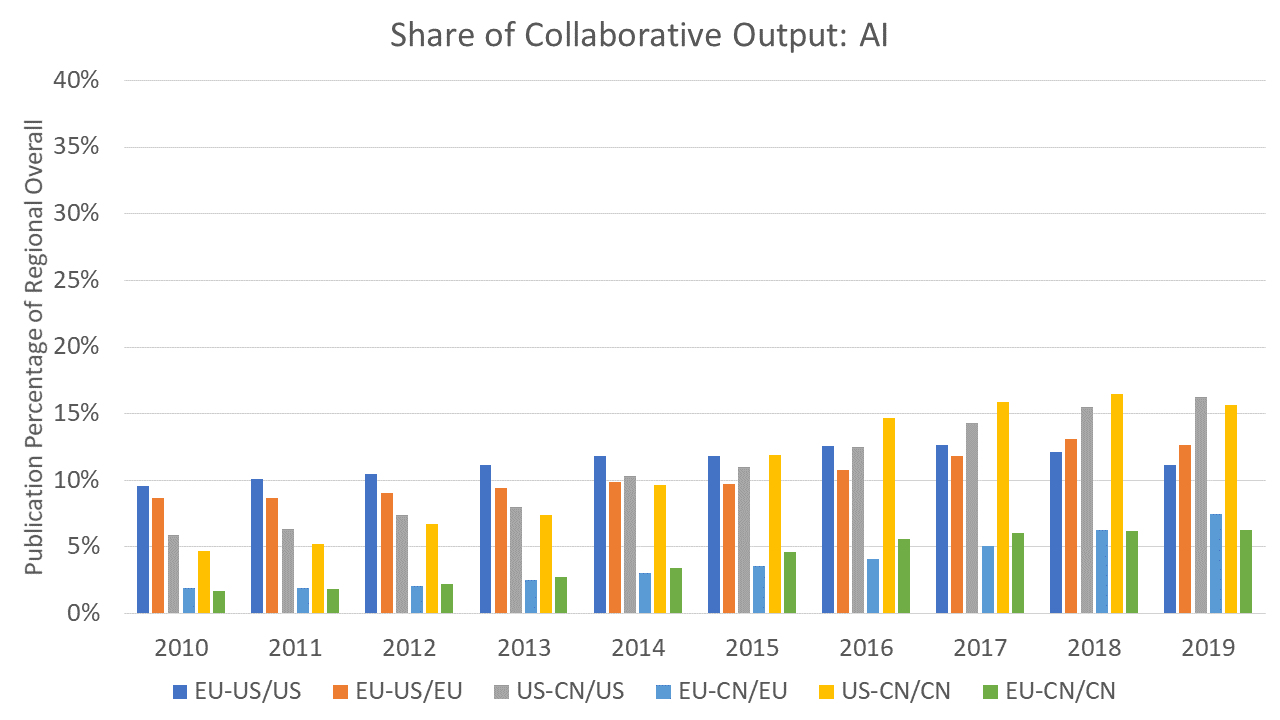} \includegraphics[width=0.49\textwidth]{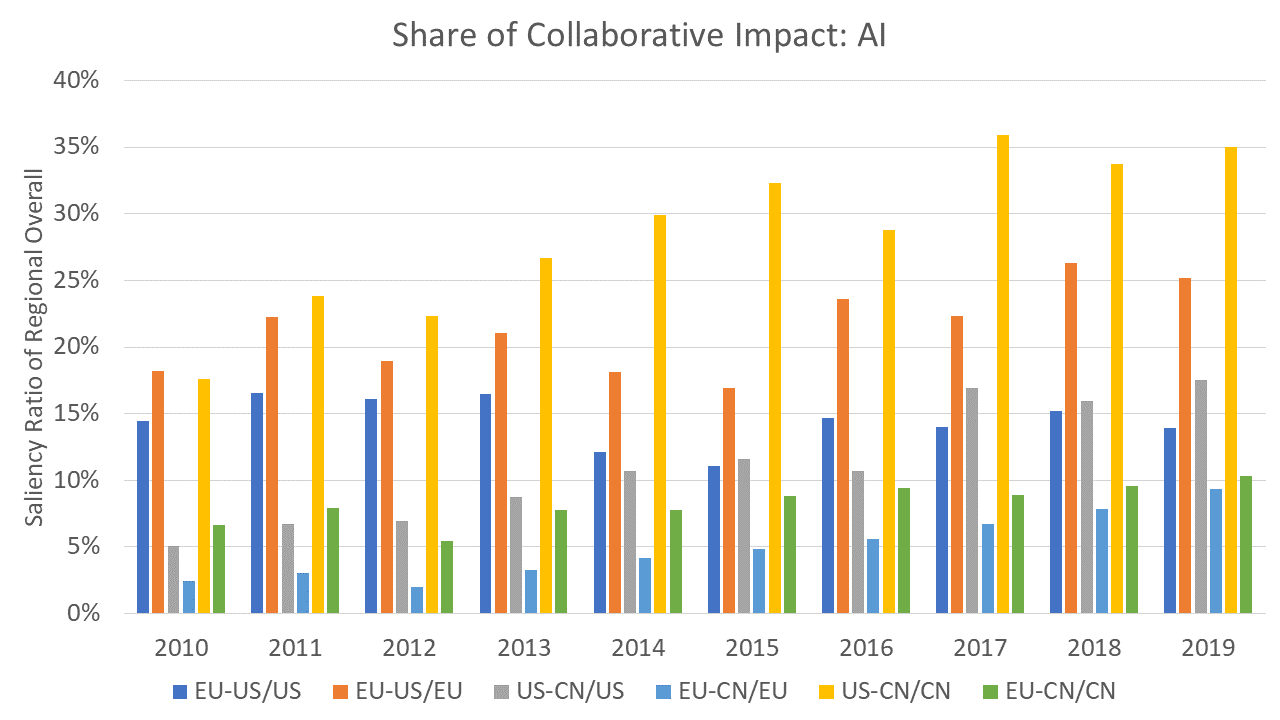}\newline
    \includegraphics[width=0.49\textwidth]{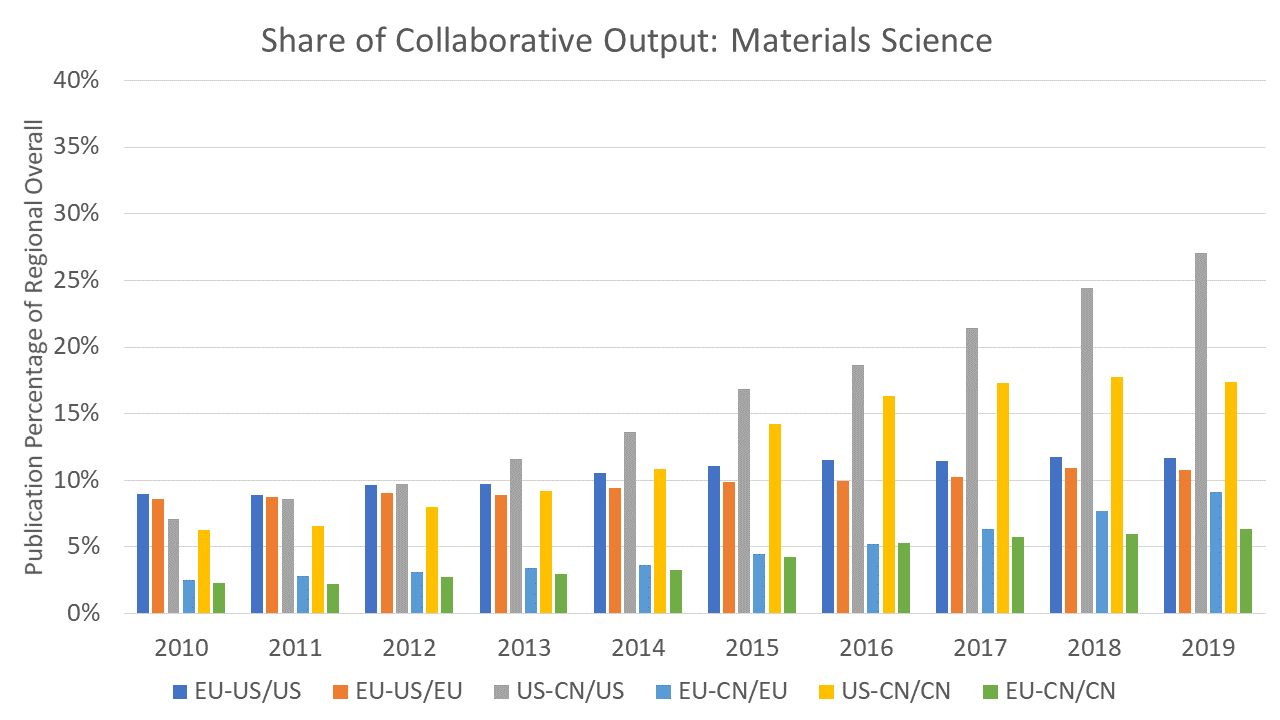} \includegraphics[width=0.49\textwidth]{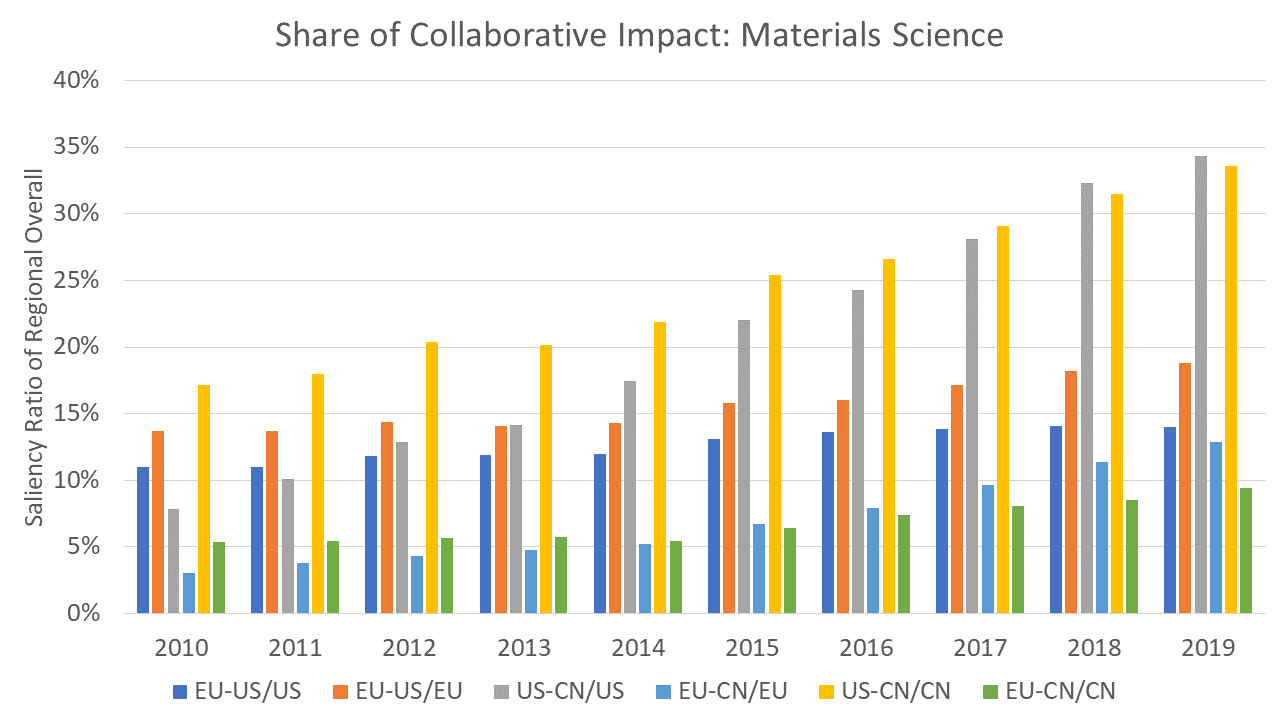}\newline
    \caption{Percentages of transregional collaboration with respective to regional totals of research output (left) and impact (right) for the fields of Medicine (top), AI (middle) and Materials Science (bottom).}
    \label{fig:CollabFields}
\end{figure}

The varying degrees of convergence in these fields reflect clearly in their respective collaboration trends as shown in Figures~\ref{fig:CountryCollabSTEM} and~\ref{fig:CollabFields} where the percentage of each country's published articles as a result of transregional collaboration (left) is juxtaposed against their corresponding impact (right) for the most recent decade. Consistently across all fields and throughout all the years are research articles with transregional collaboration among these regions increasing in their output shares, and the collaborative research account for proportionally larger impact than those that are not. Take year 2019 as an example. As shown in Figure~\ref{fig:CountryCollabSTEM}, 17.0\% of the output from China in all STEM research is a result of collaboration with US, and this transpacific research collaboration accounts for 36.0\% of the total impact from China of that year, a significant growth from 5.7\% and 20.4\%, respectively, from a decade earlier. In contrast, the US-EU collaboration has its research output hovering relatively steadily between 10.5\% and 15.4\% percentage points for either region's total, with European countries seeing their collaborative research account for a slightly larger portion of their total impact (between 20.5\% and 29.6\% versus 15.7\% to 20.3\% for the US). The fact that EU and China are gaining more in research collaboration with the historically dominant US is consistent with the economical model of convergence supported by the data both in research output and impact shown previously.

Again, the highly aggregated STEM results masquerade intricacies in its sub-fields, especially in the intimate relationship between research collaboration and convergence. Again, the results from Medicine, AI and Materials Science, representing the fields of early, middle, and advanced stages in convergence, are shown in Figure~\ref{fig:CollabFields} to illustrate this point. The field of Medicine, with research output and impact shown to be dominated by historical elites (Figures~\ref{fig:medicine}) and collaboration by transatlantic partnerships (Figures~\ref{fig:CollabOuput} and~\ref{fig:CollabImpact}), nevertheless sees significant outreach from China. Though still low in absolute quantities, the collaborations with the US and European countries are growing steadily, with their shares of impact reaching 33.6\% and 13.6\% in year 2019, out of 13.1\% and 5.4\% of China's research output, respectively. These out-sized gains in proportional impact have rivaled the European paybacks in their collaboration with the US, and can be a sustaining driver for future collaboration. 

On the other end of the research convergence is the field of Materials Science where, as shown in Figure~\ref{fig:CollabFields}, the trend has exhibited a remarkable role reversal in recent years. There, in the middle of the most recent decade, the transpacific collaboration, both in terms of research output and impact, has overtaken the transatlantic one. Most amazingly, the US-China collaboration has accounted for larger portion of US's output (27.0\% in year 2019) and impact (34.3\% in 2019) than of China's in recent years (17.3\% and 33.6\% in year 2019, respectively). Although at lower percentage points, the Sino-European collaboration tells a similar story: since the onset of the decade, such collaboration has accounted for a slightly higher portion of the research output for Europe than China, and since the mid decade, the research impact. Unlike the US, however, Europe still produces relatively more research output and impact from the collaboration with US than with China.

Straddled in the middle is the field of AI where China has surpassed US in output but still is dwarfed in impact by the US (Figure~\ref{fig:ai}), and the transpacific collaboration has lead to higher output and impact than the transatlantic (Figures~\ref{fig:CollabOuput} and~\ref{fig:CollabImpact}). These give rise to US seeing larger portions of its output and impact from its collaboration with China in recent years than with Europe earlier in the most recent decade, as shown in Figure~\ref{fig:CollabFields}. Unlike the case in Materials Science, however, the transpacific collaboration only accounts for a higher share for US than China in research output but not for the research impact yet. Specifically, transpacific collaboration has grown to account for 16.3\% of US's output versus China's 15.7\% in year 2019, while, for research impact, to 17.5\% of US's and 35.0\% of China's. In contrast, the transatlantic collaboration has steadily hovered around 10\% of US's output and 15\% of US's impact throughout the decade. The shares of Sino-European collaboration have also climbed for Europe and China in both output and impact, though the percentage points are much lower than their respective collaborations with the US.

\section{Discussion}
\label{sec:discussion}

This paper examined output, impact, and collaboration in artificial intelligence (AI) and technology research for more than 200 countries/regions in the past four decades. Using a global corpora of journal publications, conference papers, and patents that include more than 56 million articles and 1.7 billion citations, the paper examines whether developing countries are catching up (or diverging) with advanced economies in AI and scientific fields. The results show that convergence is accelerating, but only in the case of UMCs catching up with HICs, and the US becoming less dominant within HICs. The gap between UMCs and LMCs is widening, supporting evidence for divergence. Similarly, low-income countries (LICs) are diverging from LMCs and LMCs are diverging from UMCs. Various sub-fields show significantly different rates of convergence. While fields like Medicine show the least visible signs of convergence, Material Science exhibits an accelerated state of convergence. The enabling field of AI is in the middle stage of convergence, with UMCs catching up with HICs in research output but not yet in impact.

Since this study is a quantitative one, we pay special attention to the measurements of research output and impact. In addition to the broadly adopted proxies of publication counts and citation counts for output and impact, respectively, we examine the efficacy of saliency, a research impact measure included in MAG that employs a modern AI method to rectify several known problems in using citation counts. We have found that the measurement mitigates against the temporal bias in citations and its clear mathematical foundation provides a solid analytical framework to avoid the inexplicable and embarrassing conclusions derived from alternatives. The rigorous approach to quantifying research lends us confidence in describing the remarkable trends of convergence and divergence, especially the dramatic developments in recent decades. In a way, the measurements in this study are chosen rather conservatively to avoid unresolved research problems. For this reason, we have not considered data sets and software tools alone as evidence of research output or using Altmetrics for impact although these proposals have received a lot of attentions recently \cite{thelwall2013do, costas2015do, haustein2014tweeting, bornmann2014do}.

Aside from reporting observations in the data, we attempt to uncover the potential drivers of convergence by examining the heightened inter-regional collaborations, an outcome of globalization. The data clearly shows that collaborations benefits all parties involved. Specifically for the case study on the rise of China, we have found that the country's increasing output is a result of collaborating with HICs such as the US and the European countries, and these collaborations {\em always} account for more than commensurate impact. A closer look into the data suggests the rise of China is consistent with the preferential attachment process that has been theorized as a mechanism underlying the formation of naturally occurring networks, including social and research collaboration networks \cite{albert2001statistical,newman2001clustering,abbasi2012betweenness}. The central idea here is that newcomers can receive faster and broader recognition by associating themselves with those already highly recognized. Producing joint research output is certainly an effective means, and data are consistent with the catch-up effect observed for Europe catching up with the US in earlier decades, and China with the HICs more recently. The lack of similar collaboration patterns from LMCs and LICs may well be a pivotal reason to explain their divergence. Note that the preferential attachment process (or a more specialized cumulative advantage process \cite{price1976a}) does not prescribe a sufficient condition for catching up. Indeed, the theory was originally developed to explain the inequality of the Matthew effect \cite{merton1968matthew} which suggests that the highly recognized will get more recognition, and even eclipse the attribution to the newcomers if the same discoveries are made by both at the same time. Nevertheless, the data suggest the same process that leads to deepening inequality between competing parties can also be deftly employed to induce the catch-up effect through joint work.

This finding suggests that technology progress is tightly coupled with network effects, and therefore policymakers should view bilateral relationships as an opportunity rather than a constraint. Historically, the US-EU were the largest scientific collaborators. However, the US-China collaboration in AI, for instance, has become the largest AI partnership today, heralding a new wave of technology globalization. Developing countries can learn from China's growth pattern in research output and impact. The tides of ``rich getting richer'' phenomenon can be turned by building high-impact collaborations and catch up. When countries collaborate across borders, larger research impact is observed. The power of collaboration should be explicitly included when designing science policy.

Finally, the data show the rate of convergence is uneven across fields, suggesting the catch-up effect may be equally achievable. As developing countries discover the importance of collaboration in technology research, they should use network effects to their advantage to grow in output and impact in fields of their specialization. Policymakers have to make a conscious effort to exercise preferential attachment to foster collaborations with scientific leaders in fields of strategic importance.

\bibliographystyle{plain}
\bibliography{references.bib}

\end{document}